\documentclass[preprint,12pt]{elsarticle}
\usepackage{amssymb}
\journal{Astroparticle Physics}

\newcommand{\be}{\begin{equation}}
\newcommand{\ee}{\end{equation}}
\newcommand{\simless}{\lower.5ex\hbox{$\; \buildrel < \over \sim\;$}}
\newcommand{\simgreat}{\lower.5ex\hbox{$\; \buildrel > \over \sim\;$}} 
\newcommand{\mpro}{ m_{\rm p}} 
\newcommand{\mneut}{ m_{\rm n}} 
\newcommand{\mbar}{\langle{m}\rangle} 
\newcommand{\tmax}{T_{\rm max}} 
\newcommand{\tcen}{ T_{\rm{C}}}
\newcommand{\rhoc}{ \rho_{\rm{C}}}
\newcommand{\mchan}{M_{\rm Ch}} 

\newcommand{\sigvbar}{\langle\sigma v\rangle} 
\newcommand{\tcm}{T_{\rm cm}} 
\newcommand{\heiii}{{^3}{\rm He}} 
\newcommand{\hydroi}{{^1}{\rm H}} 
\newcommand{\dstar}{d}
\newcommand{\hiii}{{{}^3{\rm H}}}

\newcommand{\beryeight}{{{}^8{\rm Be}}}
\newcommand{\carbonxii}{{{}^{12}{\rm C}}}
\newcommand{\oxvi}{{{}^{16}{\rm O}}}

\begin{document}

\begin{frontmatter}

\title{{\bf On the Habitability of Universes 
without Stable Deuterium}}  

\author[a,b]{Fred C. Adams}
\author[a]{and Evan Grohs}

\address[a]{Physics Department, University of Michigan, Ann Arbor, MI 48109} 
\address[b]{Astronomy Department, University of Michigan, Ann Arbor, MI 48109} 


\begin{abstract}
In both stars and in the early universe, the production of deuterium
is the first step on the way to producing heavier nuclei.  If the
strong force were slightly weaker, then deuterium would not be stable,
and many authors have noted that nuclesynthesis would be compromised
so that helium production could not proceed through standard reaction
chains. Motivated by the possibility that other regions of space-time
could have different values for the fundamental constants, this paper
considers stellar evolution in universes without stable deuterium and
argues that such universes can remain habitable.  Even in universes
with no stellar nucleosynthesis, stars can form and will generate
energy through gravitational contraction.  Using both analytic
estimates and a state-of-the-art stellar evolution code, we show that
such stars can be sufficiently luminous and long-lived to support
life. Stars with initial masses that exceed the Chandrasekhar mass
cannot be supported by degeneracy pressure and will explode at the end
of their contraction phase. The resulting explosive nucleosynthesis
can thus provide the universe with some heavy elements. We also
explore the possibility that helium can be produced in stellar cores
through a triple-nucleon reaction that is roughly analogous to the
triple-alpha reaction that operates in our universe. Stars burning
hydrogen through this process are somewhat hotter than those in our
universe, but otherwise play the same role. Next we show that with
even trace amounts (metallicity $Z\sim10^{-10}$) of heavy elements ---
produced through the triple-nucleon process or by explosive
nucleosynthesis --- the CNO cycle can operate and allow stars to
function. Finally, we consider Big Bang Nucleosynthesis without stable
deuterium and find that only trace amounts of helium are produced,
with even smaller abundances of other nuclei.  With stars evolving
through gravitational contraction, explosive nucleosynthesis, the
triple-nucleon reaction, and the CNO cycle, universes with no stable
deuterium are thus potentially habitable, contrary to many previous
claims. 
\end{abstract}


\end{frontmatter}

\section{Introduction} 
\label{sec:intro}  

The laws of physics allow for the development of life in our universe,
but many authors have noted that sufficiently large variations could
render the universe uninhabitable
\cite{carr,bartip,hogan,reessix,aguirre,tegmark,barnes2012,schellekens}.
One partial explanation for why the laws of physics have their
observed form is that our universe is one out of many
\cite{vilenkin,guth}.  This vast collection of universes --- the
multiverse --- samples all of the possible versions of physical law.
In this scenario, the strength of the strong force could be different
in the various universes that make up the multiverse. If the strong
force were stronger, however, then diprotons and dineutrons could be
stable, and then nucleosynthesis would proceed in a different manner,
although recent work shows that stars can still operate \cite{barnes2015}.
On the other hand, if the strong force were sufficiently weaker, then
deuterium would not have a bound state. In a universe with no
deuterium, the usual stepping stone on the pathway to heavy elements
would not be available. Many authors have speculated that the absence
of stable deuterium would lead to a universe with no heavy elements at
all, and hence a lifeless universe \cite{bartip,hogan,reessix,dyson}.
The goal of this paper is to explore the possibility that stars can
provide both energy generation and nucleosynthesis, even in the
absence of stable deuterium. Under the action of gravitational
contraction alone, stars can generate enough energy, with sufficiently
long lifetimes, for a universe to be habitable. For this scenario,
massive stars will collapse at the end of their contraction phase and
provide heavy elements through explosive nucleosynthesis.  In
addition, we explore the triple-nucleon reaction, which is analogous
to the triple-alpha reaction that produces carbon in our
universe. This type of reaction provides yet another pathway for the
synthesis of heavy elements.

This type of alternate universe must still form stars in order to
operate. In our universe, however, the star formation process is
largely independent of nuclear considerations \cite{shu87,mcost}, so
that star formation could readily take place in the absence of stable
deuterium. More specifically, the interstellar medium forms objects
with a wide range of masses, and those entities have no way to tell in
advance that their final masses should be capable of achieving nuclear
fusion. Moreover, the process of star formation often forms
stellar-like bodies with masses that are too small to sustain nuclear
reactions.  These brown dwarfs are abundant in our universe, with
about one such object for every 4 or 5 ordinary stars \cite{luhman}. 

The vast majority of stars (those with masses less than about 7
$M_\odot$ \cite{palla}) are born with stellar structure configurations
that are too large in radius and too cool in central temperature to
sustain nuclear reactions. After formation, these bodies slowly
contract and are powered by the loss of gravitational potential
energy. This contraction phase ends when the stellar core becomes
sufficiently hot and dense for hydrogen fusion to take place. As a
result, nuclear fusion occurs millions of years after the formation of
most stars, defined here as hydrostatically supported objects that
have been separated from the molecular clouds that produce them.  High
mass stars ($M_\ast>7M_\odot$) follow a similar evolutionary path, but
their contraction times are shorter than their formation times. These
high mass objects are also powered by gravitational contraction in
their earliest phases, but they transition into nuclear burning
configurations before they attain their final masses.

In this paper, we assume that the star formation process proceeds as
outlined above for our universe \cite{shu87,mcost}, and consider the
subsequent evolution of stars in the absence of stable deuterium. For
this scenario, we explore four stellar processes that allow the
universe to become potentially habitable: gravitational contraction,
supernova-like explosions due to stellar collapse, the triple-nucleon
reaction (analogous to the triple-alpha reaction that produces carbon
in our universe), and finally hydrogen fusion through the CNO
cycle. The following discussion outlines our treatment of each of
these processes.

In the absence of nuclear reactions, the gravitational contraction
phase described above for pre-main-sequence stars in our universe will
continue over much longer spans of time. This paper shows that stars
can generate enough energy, over sufficiently long timescales, to
sustain life. In this scenario, small stars will contract until their
central regions become degenerate, and their luminosities will slowly
fade.  For high mass stars, however, degeneracy pressure is not
sufficient to support the mass of the star and it will experience
catastrophic collapse. The subsequent implosion compresses the stellar
core to enormous densities and temperatures so that explosive
nucleosynthesis can take place, even in the absence of stable
deuterium nuclei. The result is much like Type Ia supernovae in our
universe, where these explosions are produced by the collapse of white
dwarfs. As a result, low mass stars can provide energy through
gravitational contraction, whereas high mass stars can provide
nucleosynthesis through collapse.

The mass scale that marks the boundary between high mass stars and low
mass stars is the Chandrasekhar mass \citep{chandra}.  This scale
represents the largest mass that can be supported by non-relativistic
degeneracy pressure of electrons \citep{clayton,kippenhahn,phil} and
depends on the chemical composition of the object.  For stellar
evolution in our universe, the Chandrasekhar mass $\mchan$ is usually
evaluated under the assumption that $A/Z=2$, where $Z$ is the atomic
number and $A$ is the atomic weight (because stars that become white
dwarfs in our universe are mostly made of carbon and oxygen). For the
first generation of stars in a universe without stable deuterium, the
composition is expected to be pure hydrogen so that $A/Z=1$. Since
$\mchan\propto(A/Z)^{-2}$ \citep{chandra,clayton,kippenhahn}, the
Chandrasekhar mass will be larger than in our universe, namely
$\mchan\approx5.6M_\odot$.

Stars can also burn hydrogen through other reaction chains that do not
rely on the existence of stable deuterium. We first consider the
triple-nucleon reaction, which is analogous to the triple-alpha
reaction that produces carbon in our universe. In this latter case,
the $^8$Be isotope is unstable, so the simplest reaction $^4$He +
$^8$Be $\to$ $^{12}$C is suppressed. In spite of being unstable,
$^8$Be nuclei are produced in stellar cores due to nuclear statistical
equilibrium (NSE).  The production of $^8$Be is not energetically
favored because the isotope is unstable, so that the standing
population is small.  Nonetheless, enough $^8$Be exists so that carbon
can be produced.  In a roughly similar scenario, deuterium nuclei will
be produced in stellar cores, even though they are unstable and not
energetically favored. Because of the inefficiency of the weak
interaction, the stellar core will not generally reach NSE, so that
the abundances of nuclear species must be calculated more explicitly.
However, the resulting population of deuterium nuclei can interact
with protons to form $^3$He, which eventually fuses into $^4$He. The
abundance of deuterium is a steeply increasing function of
temperature, so that this reaction requires stellar cores to be hotter
and denser than their counterparts in hydrogen burning stars in our
universe.

The final process that we consider is the Carbon-Nitrogen-Oxygen (CNO)
cycle. In this chain of reactions, carbon is used as catalyst to fuse
hydrogen into helium \cite{clayton,kippenhahn}.  Although no deuterium
is required for this reaction chain, at least some carbon must be
present. As a result, some previous epoch of nucleosynthesis, through
collapse of high mass degenerate stars and/or through the triple
nucleon reaction, must occur in order for this channel to be viable.
This paper shows that the CNO cycle can operate --- and drive stellar
evolution much like that in our universe --- provided that the stellar
metallicity $Z>10^{-10}$. For comparison, the metallicity of the Sun
$Z_\odot\approx0.013$ \cite{asplund} and the metallicity of the most
metal-poor stars observed in the universe have at least $Z\sim10^{-7}$
\cite{frebel}.

For the scenario considered here, with no stable deuterium, the
universe is expected to emerge from its early epochs with essentially
no elements heavier than hydrogen. During the first several minutes of
cosmic time, during the epoch known as Big Bang Nucleosynthesis (BBN),
our universe converts about one fourth of its mass into helium, with
trace amounts of deuterium and lithium \cite{kolbturner,kawano}.
Without stable deuterium as an intermediate state, this paper shows
that BBN produces only traces amounts of helium, with mass fraction
$X_4\sim10^{-14}$. We can thus assume that the first generation of
stars have an almost pure hydrogen composition.  Compared to the
conditions realized during BBN, however, stellar cores can provide
higher temperatures, higher densities, and longer time scales for
confinement. These properties thus allow stars to achieve hydrogen
fusion, even in the absence of stable deuterium. 

This paper is organized as follows. Stellar evolution through
gravitational contraction alone, with no nuclear burning, is
considered in Section \ref{sec:gravity}. The largest stars must
collapse at the end of their contraction phase and can potentially
produce heavier elements through explosive nucleosynthesis. The
triple-nucleon process is considered in Section \ref{sec:ppp}. After
showing that stellar cores generally do not have time to reach NSE, we
develop a generalized reaction network that keeps track of free
neutrons. These reactions allow stars to generate energy and evolve
much like stars in our universe, albeit with higher temperatures both
in the core and on the surface. Next we show that stars with trace
amounts of carbon can operate through the CNO cycle and this chain of
nuclear reactions is explored in Section \ref{sec:cno}. For
completeness, we revisit the epoch of Big Bang Nucleosynthesis in
Section \ref{sec:bbn} and show that BBN produces essentially no heavy
nuclei. The paper concludes, in Section \ref{sec:conclude}, with a
summary of our results and a discussion of their implications.

\section{Stellar Evolution through Gravitational Contraction} 
\label{sec:gravity} 

In this section we consider the evolution of stars in the absence of
any nuclear reactions. These stars will evolve through gravitational
contraction (only) and can provide strategically situated planets with
an ample supply of energy. In order to substantiate this claim, we use
both analytic arguments (Section \ref{sec:gcontract}) and numerical
simulations from the {\sl\small MESA} computational package (Section
\ref{sec:mesa}).  In this scenario --- with no sustained nuclear
burning --- the synthesis of heavy elements can take place during the
collapse that marks the end of evolution for sufficiently massive
stars (Section \ref{sec:explode}). Note that in this section we retain
the constants $(k,c,\hbar)$, consistent with most literature on
stellar structure.  For the remainder of the paper, however, we set
these constants to unity and work in natural units.

\subsection{Gravitational Contraction of Non-burning Stars}
\label{sec:gcontract}  

Using the standard arguments with homology relations for the equations
of stellar structure (see chapter 20 of \cite{kippenhahn}), we can
find scaling relations for the luminosity as a function of stellar
properties. For radiative stars using Kramer's opacity law
$\kappa\propto\rho T^{-7/2}$ \cite{clayton,kippenhahn,phil}, where
$\rho$ is density and $T$ is temperature, one finds 
\be
L_\ast \sim L_0 \left({ M_\ast \over M_0} \right)^{11/2} 
\left({ R_\ast \over R_0} \right)^{-1/2} 
\left( {\mu G \over \mu_0 G_0} \right)^{15/2} \,, 
\label{homosun} 
\ee
where $L_\ast$, $M_\ast$, $R_\ast$, and $\mu$ represent the stellar
luminosity, mass, radius, and mean molecular weight. The subscript
`0' denotes reference values.  For more massive stars that ionize their
interiors, the stellar opacity is given by the Thompson cross section,
and the scaling relation simplifies to the form 
\be
L_\ast \sim L_0 \left({ M_\ast \over M_0} \right)^3 \,. 
\label{homobig} 
\ee 

Consider a newly formed star with no nuclear reactions and a
relatively large mass $M_\ast > 1 M_\odot$. At first, the star will be
powered by gravitational contraction, just like pre-main-sequence
stars operate in our universe. The gravitational energy is given by 
\be
U = - f_1 {GM_\ast^2 \over R_\ast} \,, 
\ee
where $f_1$ is a dimensionless constant of order unity, 
so that the stellar luminosity is given by 
\be
L_\ast = {dE \over dt} = -f_1 {GM_\ast^2 \over R_\ast^2} 
{dR_\ast \over dt} \,. 
\label{drdt} 
\ee
As the star contracts, its central temperature and central 
density grow larger according to the scaling laws 
\be
\tcen \propto {1 \over R_\ast} \qquad {\rm and} \qquad 
\rhoc \propto {1 \over R_\ast^3} \,.  
\ee
The central temperature is thus given by an expression 
of the form 
\be
k\tcen = f_2 {GM_\ast \mpro \over R_\ast} \,,
\ee
where $f_2$ is another dimensionless constant of order unity, and
$\mpro$ is the proton mass. The derivation of this relation assumes
that the pressure is provided by the ideal gas law. This assumption
breaks down under two conditions: [a] the star becomes too dense, so
that degeneracy pressure dominates, and [b] the star is too massive,
so that the pressure required to support the star leads to a
temperature large enough that radiation pressure dominates.  

For sufficiently massive stars, the central temperature thus has a
maximum value given by the transition to an equation of state
dominated by radiation pressure. This maximum value has been 
derived previously \cite{adams2008,adamsnew} and has the form 
\be
k \tmax \approx 1.4 \left( {m_{\rm ion} \over \mbar} \right)^{8/3} 
m_e c^2 \approx 4.5 {\rm MeV}\,,  
\label{tmaxbig} 
\ee
where $m_{\rm ion}$ is the mass of the nuclei and $\mbar$ is the mean
mass per particle. Since we are considering massive stars with a pure
hydrogen composition, $m_{\rm ion}=\mpro$ and $\mbar\approx\mpro/2$ 
(assuming efficient ionization of the stellar interior). Note that
electrons become relativistic at this energy scale.  When electrons
are relativistic, they produce less degeneracy pressure for a given
density, and cannot support a star against catastrophic collapse.
Stars with masses above the Chandrasekhar limit will eventually 
collapse and achieve this maximum temperature. 

For stars with smaller masses ($M_\ast<\mchan$), the onset of
degeneracy enforces a maximum temperature in the stellar core. This
maximum value has been derived previously \cite{clayton,phil,adams2008} 
and can be written in the form
\be
k\tmax = {5 \over 36 (4\pi)^{2/3}} 
{G^2 M_\ast^{4/3} \mpro^{8/3} m_e  \over \hbar^2 } \approx 
5.8 \,\,{\rm keV} \,\, \left({M_\ast \over M_\odot}\right)^{4/3}\,.
\label{tmaxdegen} 
\ee
The largest temperature that can be attained without the collapse 
of the stellar core is given by equation (\ref{tmaxdegen}) for the 
Chandrasekhar mass. This temperature is $\sim60$ keV or $\sim$ 
$7\times10^8$ K. 

The contraction time is determined by the integration of equation
(\ref{drdt}). As shown above, the luminosity is nearly constant as the
radius shrinks for sufficiently massive stars. The time required for
the star to contract to a radius $R_\ast$ is thus given by 
\be
\Delta t = f_3 {G M_\ast^2 \over L_\ast R_\ast} \,,
\ee
where $f_3$ is a dimensionless parameter of order unity. Since the
star has a maximum temperature, it must have a minimum radius given by 
\be
R_{\rm min} = f_2 {GM_\ast \mpro \over k \tmax} \,. 
\ee
Combining these equations we thus obtain 
\be
\Delta t = {f_3 \over f_2} {M_\ast \over \mpro} 
{k \tmax \over L_\ast} \,.
\label{radtime} 
\ee
For $M_\ast=10M_\odot$, the luminosity $L_\ast\approx10^4L_\odot$, and
the total evolution time becomes $\Delta t \approx 10^7$ yr. For stars
with much larger masses, $M_\ast \gg 10 M_\odot$, the luminosity
scales as $L_\ast\sim M_\ast$, so that the time required to reach the
maximum temperature is nearly the same for all massive stars. For
stars in the mass range $\mchan<M_\ast<10 M_\odot$, the stellar
luminosity scales according to $L_\ast\sim{M_\ast}^3$, so that the
total evolution time $\Delta t \sim M_\ast^{-2}$. Decreasing the
stellar mass to $\mchan$ thus provides (only) a factor of three
increase in $\Delta t$. As a result, all massive stars (with 
$M_\ast>\mchan$) have evolution time scales of order 10 Myr.

For stars less massive than the Chandrasekhar mass $(\sim5.6M_\odot)$,
the central regions never reach the point where electrons are
relativistic. Instead they reach the maximum temperature given by
equation (\ref{tmaxdegen}). For stellar masses well below the
Chandrasekhar mass, we can use this result to evaluate the time scale
of equation (\ref{radtime}), which leads to the result 
\be
\Delta t \approx 10^7\,\,{\rm yr}\, 
\left({M_\ast \over M_\odot}\right)^{-2/3} \,, 
\ee
where we have used the luminosity scaling $L_\ast\sim M_\ast^3$. The
evolution time is thus slowly varying.  For sufficiently small stars,
however, the interiors are not fully ionized, so that we must use the 
scaling relation of equation (\ref{homosun}) instead of that of
equation (\ref{homobig}). With the steeper scaling relation 
$L_\ast \sim M_\ast^{11/2}$, the evolution time becomes longer. 

\begin{figure}
\begin{center}
\includegraphics[scale=0.67]{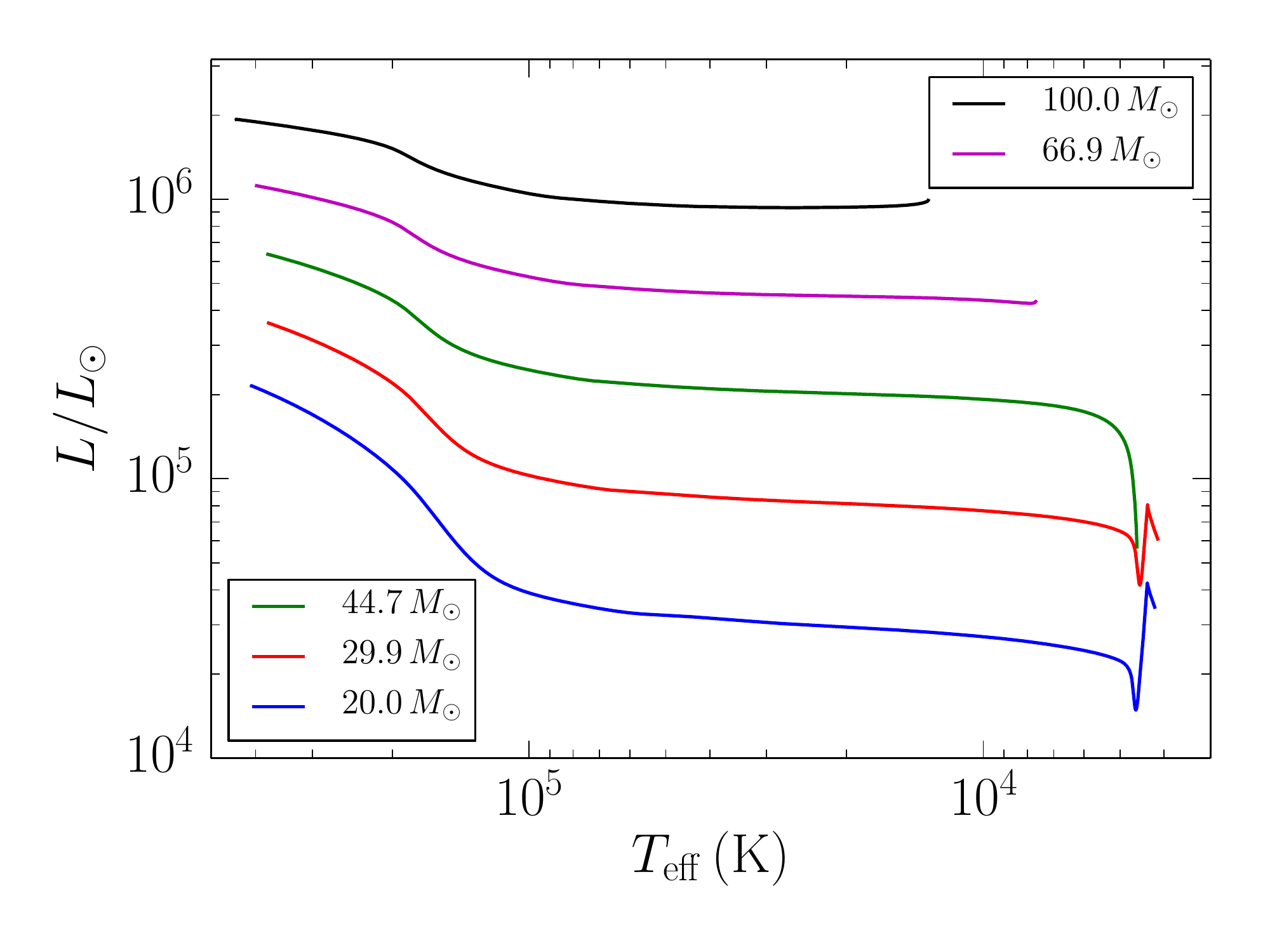}
\end{center}
\caption{H-R diagram for stars in universes without stellar 
nucleosynthesis. The tracks illustrate the evolution of high 
mass stars with masses from $M_\ast=20$ to 100 $M_\odot$
(mass increases from bottom to top). } 
\label{fig:hrone} 
\end{figure}

\begin{figure}
\begin{center}
\includegraphics[scale=0.67]{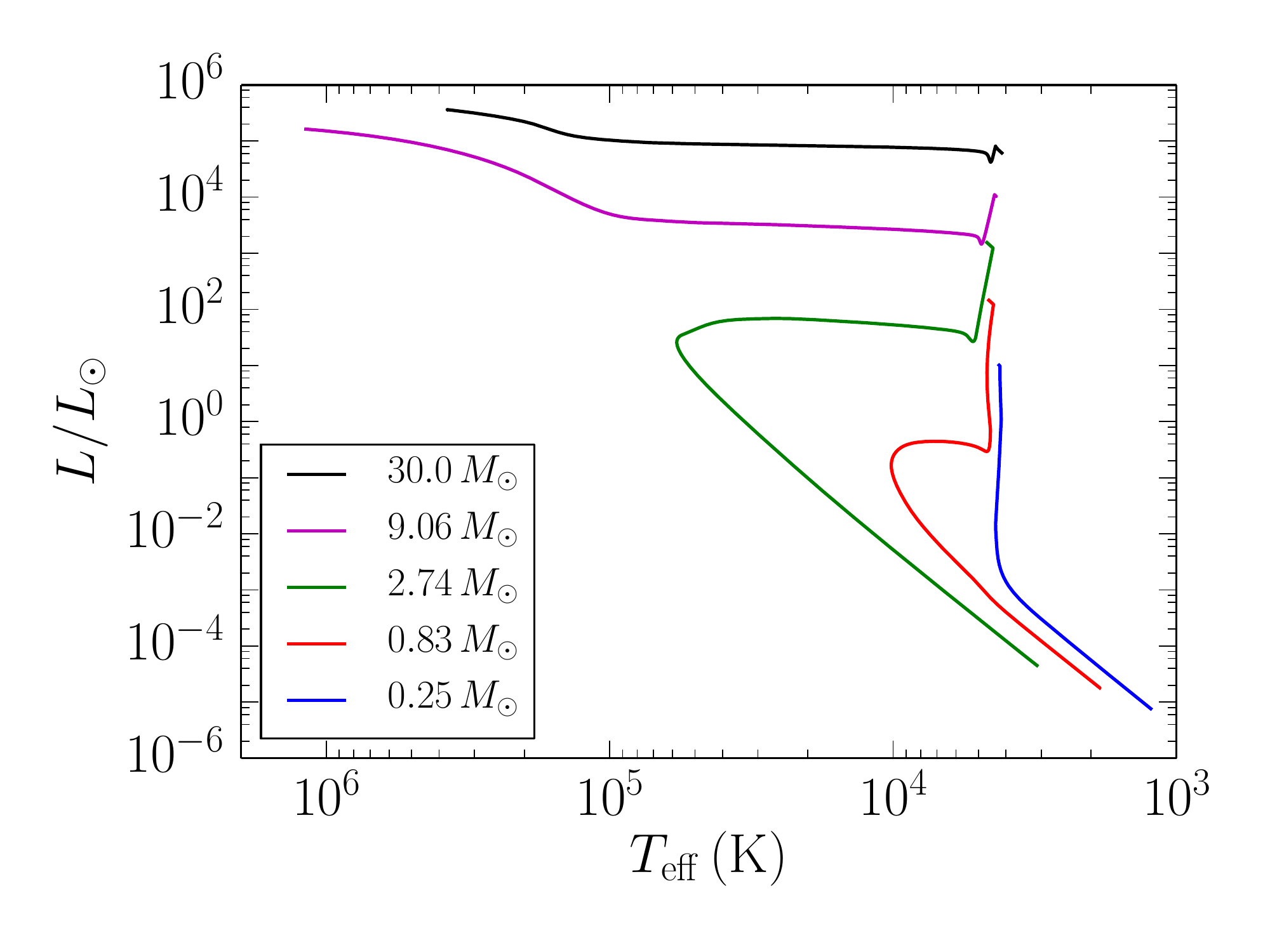}
\end{center}
\caption{H-R diagram for stars in universes without stellar 
nucleosynthesis. The tracks illustrate the evolution of stars
with varying masses from $M_\ast$ = 0.25 $M_\odot$ (right-most
blue track) to 30 $M_\odot$ (upper-most black track). }
\label{fig:hrtwo} 
\end{figure}

\subsection{Stellar Evolution Simulations without Nuclear Reactions} 
\label{sec:mesa} 

To illustrate the possible types of stellar evolution that can take
place in universes with no deuterium, we first consider the case where
no nucleosynthesis takes place in stars. Using the state-of-the-art
computational package {\sl\small MESA} \cite{paxton,paxtonb}, we
evolve a collection of stars under the action of gravitational
contraction only. The stars are assumed to start in a configuration
comparable to the initial states for stars in our universe, i.e., with
radii several times larger than their main-sequence sizes and
correspondingly lower central temperatures \cite{palla}. The stars are
also assumed to have zero metallicity with a pure hydrogen
composition. The stars then evolve via gravitational contraction.

The evolution of these stars in the H-R diagram is illustrated in
Figures \ref{fig:hrone} and \ref{fig:hrtwo}. Evolutionary tracks are
shown for a wide range of stellar masses, where $M_\ast$ =
$0.25-100M_\odot$.  Figure \ref{fig:hrone} shows the tracks for the
upper end of the mass range (note that the luminosity scale is much
smaller than that of Figure \ref{fig:hrtwo}). In all cases, the stars
begin with a nearly vertical track on the right side of the H-R
diagram. During this early phase of evolution, the stellar interiors
are convective, and gravitational contraction takes place at nearly
constant effective temperature. For the high mass stars, this phase is
short-lived ($\Delta t \sim 10^4$ yr), whereas low mass stars remain
convective much longer ($\Delta{t}>10^7$ yr). The process of star
formation itself takes $\sim10^5$ years, so that stars are not
optically visible for the first part of the evolution shown here.
Nonetheless, stars in our universe experience an analogous convective
phase (except for massive stars with $M_\ast>7M_\odot$, which evolve
through their convective phase while they are still in the process of
forming). 

Following the convective phase, the stellar interiors become radiative
and the tracks in the H-R diagram become nearly horizontal,
corresponding to nearly constant luminosity (see equation
[\ref{homobig}]). In our universe, stars reach a central temperature
where hydrogen fusion takes place and gravitational contraction halts.
In the case considered here, with no nuclear reactions, the
contraction continues.  This relatively long-lived phase of constant
luminosity is conducive to supporting habitable planets. The lifetime
of this phase is given approximately by equation (\ref{radtime}) for
stars that are massive enough to remain radiative and have hot enough
interiors so that the stellar opacity is given by the Thompson cross
section. These conditions require the stars to be at least as massive
as the Sun. For smaller stars, the interior is not fully ionized, and
the opacity is given by Kramer's law. Under these conditions, where
the luminosity is given by equation (\ref{homosun}), the time spent on
the horizontal evolutionary track becomes shorter. For stars less
massive than $\sim0.5M_\odot$, the stellar interior tends to stay
convective for most of the stellar lifetime. These smaller stars tend
to evolve directly from their convective tracks to their degenerate
tracks. As a result, although small stars have long total lifetimes,
the time spent in the radiative phase, with relatively constant
luminosity, is shorter. 

\begin{figure}
\begin{center}
\includegraphics[scale=0.67]{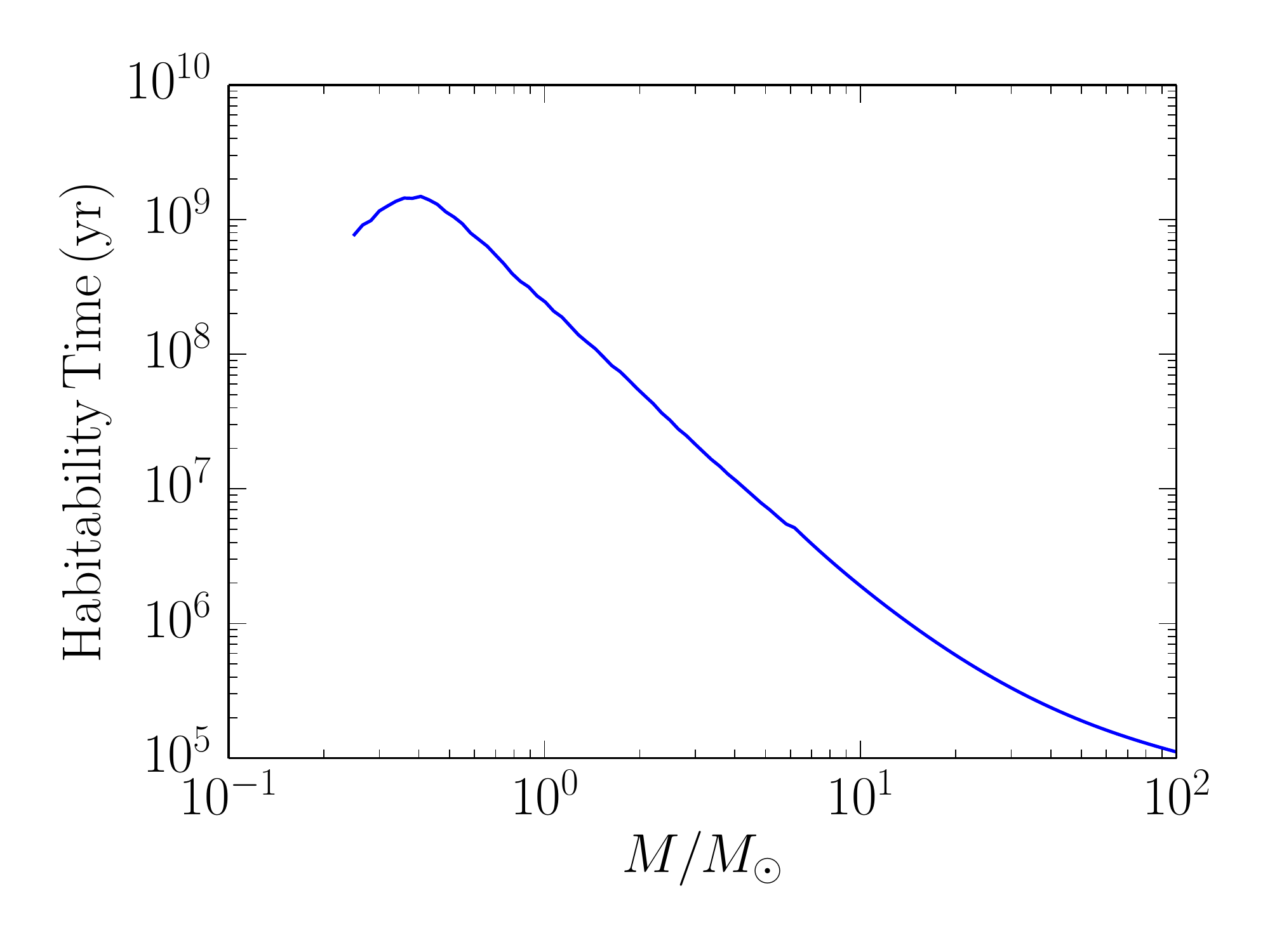}
\end{center}
\caption{Effective time for habitability for stars powered by 
gravitational contraction only. The lifetime shown here corresponds 
to the time before the stars become degenerate. High mass stars 
quickly collapse after this point in evolution, whereas low mass 
stars slowly fade and follow white-dwarf-like tracks in the 
H-R diagram (see Figure \ref{fig:hrtwo}). } 
\label{fig:habtime} 
\end{figure}

To illustrate the trends described above, Figure \ref{fig:habtime}
plots the effective habitability time for stars as a function of their
stellar mass. This habitability time is defined differently for high
mass and low mass stars. For low mass stars, the stellar core reaches
a maximum temperature when the star becomes dense enough to become
dominated by degeneracy pressure. At this point in its evolution, the
stellar luminosity decreases and the track in the H-R diagram follows
the trend appropriate for white dwarfs in our universe. This turning
point thus marks the end of the phase with nearly constant
luminosity. We also note that the time spent in the early convective
phase (depicted by the nearly vertical tracks in the H-R diagram) is
short compared to the total. Larger stars have too much mass to be
supported by degeneracy pressure. These stars also reach central
temperatures and densities where electrons are degenerate.  In these
stars, however, the electrons are relativistic and cannot provide
enough pressure to hold up the star. At this point in evolution, the
high mass stars transition from gradual gravitational contraction to
rapid gravitational collapse. These latter stages of collapse are
short, so that the lifetime shown in Figure \ref{fig:habtime}
represents the total evolutionary time of the star. The boundary
between the two types of behavior described here occurs at the
Chandrasekhar mass, where $\mchan\approx5.6M_\odot$ for stars composed
of pure hydrogen. Notice, however, that the habitability time, as 
plotted in Figure \ref{fig:habtime}, does not show a sharp boundary 
at the Chandrasekhar mass. Stars with masses somewhat less than 
$\mchan$ display the scaling law $\Delta t \propto M_\ast^{-2}$ 
expected for stars with $M_\ast>\mchan$. 

\begin{figure}
\begin{center}
\includegraphics[scale=0.67]{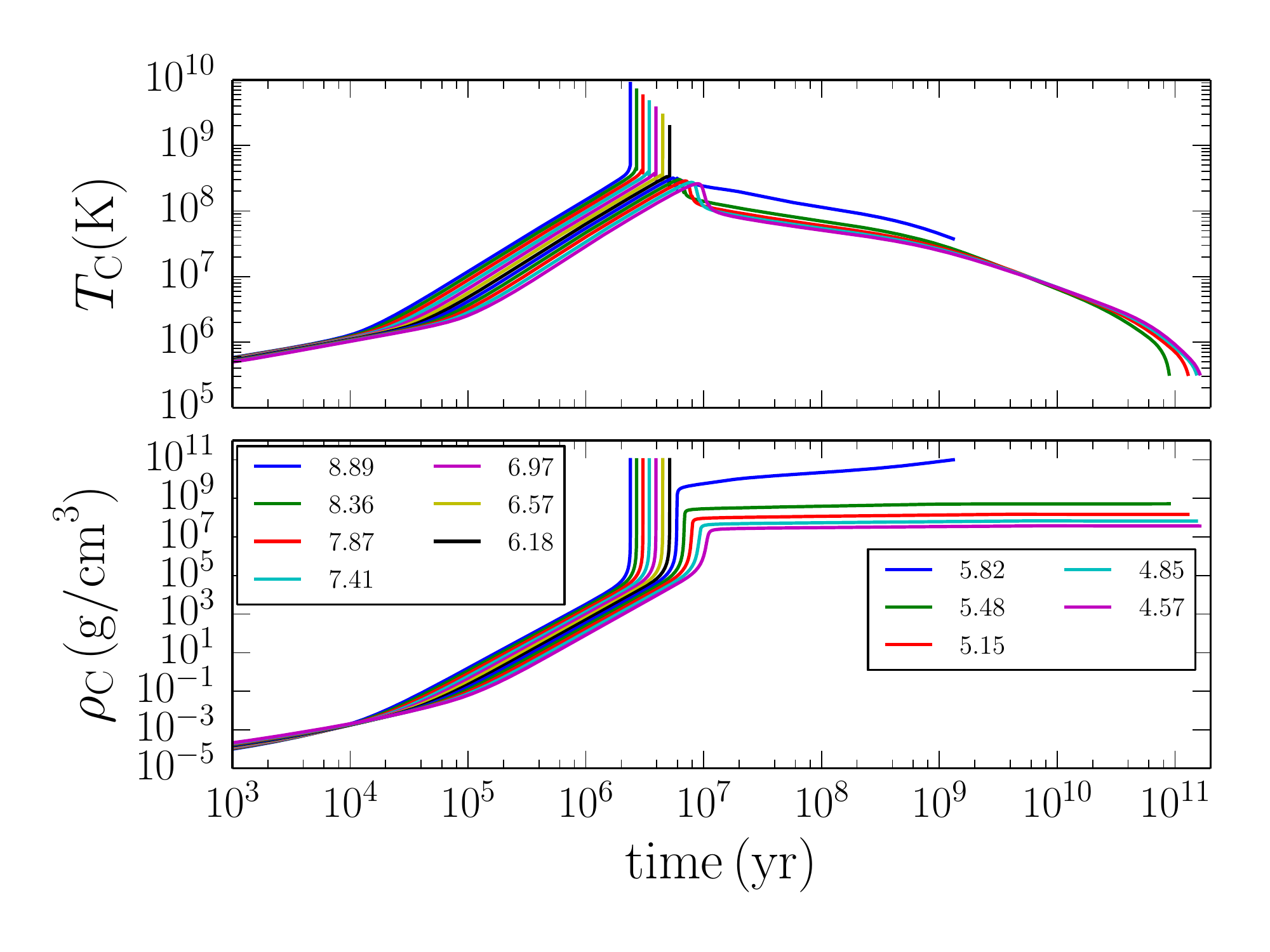}
\end{center}
\caption{Central temperature and central density as a function of time  
for stars powered by gravitational contraction only. The various
curves correspond to a range of masses, as labeled (in units of solar
masses), both above and below the Chandrasekhar mass (where
$\mchan=5.6M_\odot$ for stars with a pure hydrogen composition).  }
\label{fig:temptime} 
\end{figure}

Figures \ref{fig:hrtwo} and \ref{fig:habtime} show that stars can
provide substantial luminosity, at a relatively constant value, over
time scales longer than 1 Gyr. For a star with mass $M_\ast$ = 0.8
$M_\odot$, for example, the luminosity due to gravitational
contraction is just under 1 $L_\odot$ over a time of just under 1 Gyr.
The total evolutionary times are even longer, although the
luminosities fall with time. These considerations suggest that stars
powered by gravitational contraction are sufficient as the required
energy sources for habitability. In order for life to develop,
however, some process must also drive nucleosynthesis. We also note 
that the luminosities of these stars are not as constant as those 
for hydrogen-burning stars in our universe. It would be useful to 
know how much variability can be tolerated and still allow for 
habitability. 

Figure \ref{fig:temptime} shows the central temperature and density
for a collection of stars with a range of masses. We note that the
evolution is significantly different for stars above and below the
Chandrasekhar mass, where $\mchan\approx5.6M_\odot$ for pure
hydrogen.  For stars with masses $M_\ast < \mchan$, the central
temperature rises steadily for $\sim10$ Myr and then reaches a maximum
value of order $\tmax\sim10^8$ K. This maximum temperature
varies with stellar mass and is given by equation (\ref{tmaxdegen}) to
good approximation.  For the largest mass, $M_\ast=\mchan$, the
maximum temperature $\tmax\approx7\times10^8$ K.  At this
time, the central density reaches a corresponding maximum value of
order $\rhoc\sim10^9$ g cm$^{-3}$. At this evolutionary stage,
the star is supported by the non-relativistic degeneracy pressure of
its electrons.  At later times, the central temperature falls as the
stellar core cools, whereas the central density remains nearly
constant. The cooling time for the degenerate objects is relatively
long (up to a {\sl several} Gyr \cite{wood}), which allows the stars
to remain relatively luminous and leads to the lifetimes shown in
Figure \ref{fig:habtime}.

For more massive stars, $M_\ast>\mchan$, the time development of the
central temperatures and densities follow trajectories similar to
those of low mass stars up to times $t\sim$ {\sl few} Myr. Instead of
being supported by degeneracy pressure at this stage, however, these
heavier stars transition from a state of relatively slow contraction
to more rapid collapse. Both the central temperature and density
increase sharply. The curves in Figure \ref{fig:temptime} become
nearly vertical, indicating that collapse takes place quickly, with
extremely little additional time elapsed. These stars are expected to
collapse until they reach extreme densities so that they either become
black holes and/or experience explosive nucleosynthesis, analogous to
collapsing degenerate objects in our universe. For the evolutionary
tracks presented here, however, the {\sl\small MESA} code is unable to
follow stars after the densities exceed $\rhoc\sim10^{11}$ g
cm$^{-3}$. At this stage, the central temperatures are of order $\tcen
\sim 10^{10}$ K $\approx 0.86$ MeV, comparable to the mass difference
between the proton and neutron (1.29 MeV). At these enormous
temperatures, weak interactions (e.g., $e^{-} + p \to n + \nu_e$) are
no longer suppressed and nuclear reactions take place readily. We 
thus expect nucleosynthesis to occur efficiently during the collapse 
of these stars, even in the absence of stable deuterium. 

Figure \ref{fig:temptime} shows that the time evolution of central
temperature and density proceeds differently for high mass stars and
low mass stars. However, the boundary between stars that collapse and
those that are supported by degeneracy pressure does not seem to occur
at the quoted value of the Chandrasekhar mass $\mchan=5.6M_\odot$.
Instead, stars of slightly larger mass (e.g., $M_\ast=5.82M_\odot$ in
the Figure) do not collapse in the simulations. The Chandrasekhar mass
is derived assuming that the stellar structure is given exactly by an
$n=3/2$ polytrope \cite{chandra}, which applies in the limit where the
stellar material is completely degenerate and has zero temperature. 
In contrast, the stars in the simulations depicted in Figure 
\ref{fig:temptime} are not pure polytropes. This departure delays
their collapse. If the simulations were continued to much longer
times, the stellar material would eventually cool enough to become
fully degenerate.  In that limit, all stars with masses
$M_\ast>\mchan$ are expected to collapse.

\begin{figure}
\begin{center}
\includegraphics[scale=0.67]{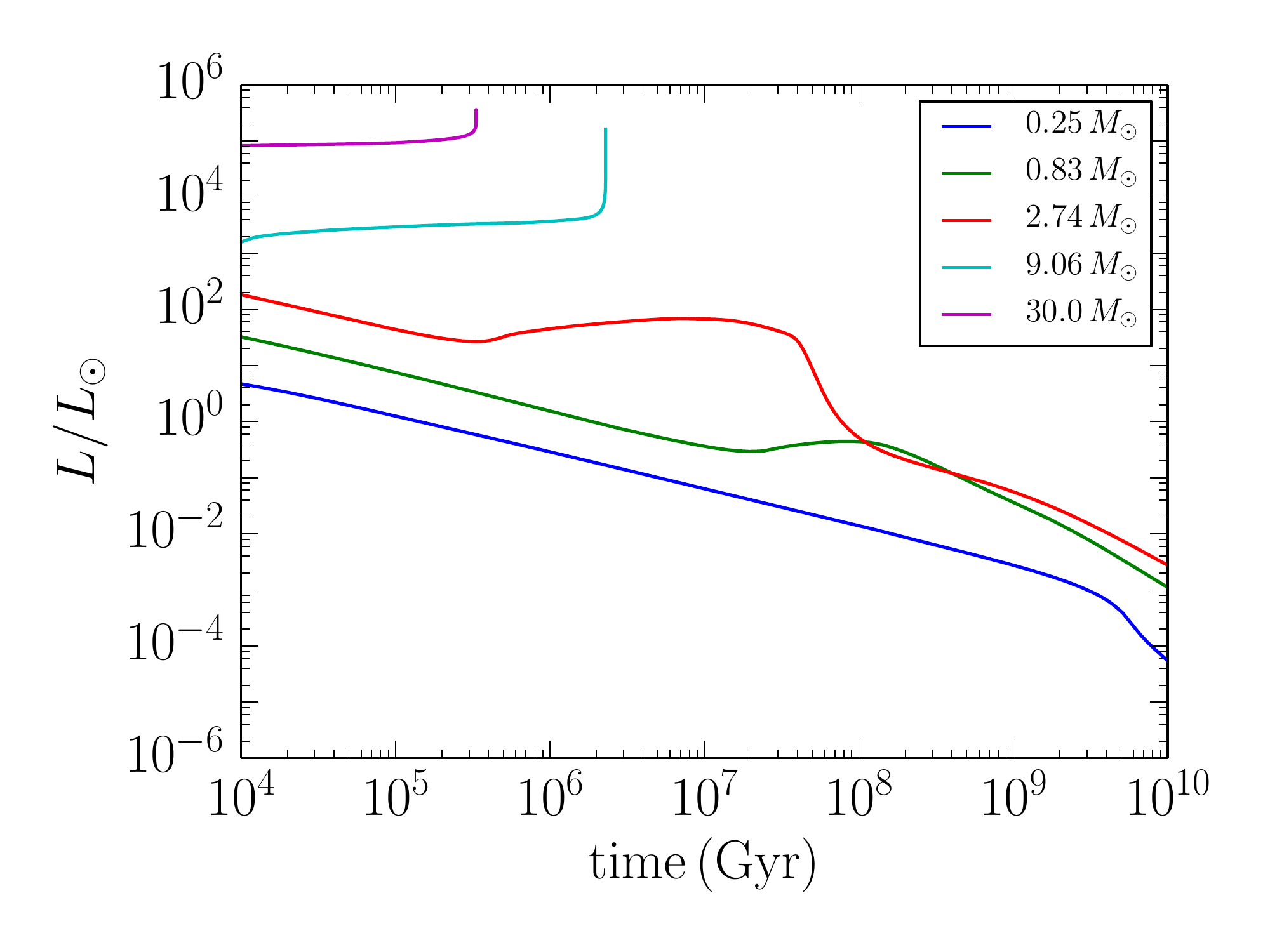}
\end{center}
\caption{Luminosity as a function of time for stars powered by  
gravitational contraction only. The various curves correspond to a
range of masses, as labeled, both above and below the Chandrasekhar
mass. }
\label{fig:lumtime} 
\end{figure}

The stellar luminosity is plotted as a function of time in Figure
\ref{fig:lumtime}. The stars show three different types of behavior.
For high mass stars, above the Chandrasekhar limit, the luminosity is
roughly constant with time until the central regions become both
relativistic and degenerate. At this point, the stars experience core
collapse and the luminosity spikes upward. The total evolutionary time
for these massive stars is short, $\Delta{t}<10^7$ yr.  For the
smallest stars, roughly those with mass $M_\ast<0.5M_\odot$, the
luminosity slowly decreases over the entire stellar lifetime. These
small stars are convective over their entire lifespan, until they
become degenerate, and have no radiative phase. For stars with
intermediate masses, roughly in the range $0.5M_\odot<M_\ast<\mchan$,
the luminosity decreases during an early convective phase, but then
goes through a radiative phase with slowly varying luminosity. After
the stars become degenerate, their luminosity decreases again. 

The corresponding tracks in the H-R diagram for these stars are shown
in Figure \ref{fig:hrtwo}, which provides the surface temperatures of
the stars in addition to their luminosity. Over the phases of nearly
constant luminosity for stars with intermediate masses, the surface
temperature steadily increases (see the nearly horizontal tracks in
Figure \ref{fig:hrtwo}). For a given stellar luminosity $L_\ast$ and
surface temperature $T_\ast$, the stellar radius $R_\ast$ can be
determined through the usual photospheric outer boundary condition
$L_\ast=4\pi R_\ast^2\sigma T_\ast^4$. The stellar radius thus
decreases while the surface temperature increases, such that the
luminosity remains nearly constant. The tracks in the H-R diagram show
that the surface temperature reaches a maximum value and then
decreases. After this point, the stellar luminosity, temperature, and
radius all decrease with time. 

Figure \ref{fig:lumtime} indicates that the largest stars with
$M_\ast>\mchan$ have short lifetimes ($<10$ Myr) and are not good
candidates to host potentially habitable planets. The smallest stars
with $M_\ast<0.5M_\odot$ have sufficiently long lifetimes, but exhibit
steadily decreasing luminosity. As a result, the most promising stars
for hosting planets are those with intermediate masses.  These stars
have substantial luminosity over relatively long spans of time. For
example, stars with mass $M_\ast=1-3M_\odot$ have luminosity
$L_\ast>0.07L_\odot$ over time scales longer than $\sim1$ Gyr. The
luminosity subsequently decreases steadily after this epoch. The onset
of steadily declining power can thus be used to mark the end of the
habitable phase, although this boundary is not sharp.  Even these
latter stars are not ideal --- although the lifetimes are long and the
luminosities are large, the luminosities are not as steady as those in
our universe. As a point of comparison, the faint early Sun had a
luminosity that was only smaller by $\sim25\%$, but that change might
have led to Earth freezing over. The variations in luminosity for the
stars considered in this section vary to a greater degreee and thus
make habitability more difficult. 

\subsection{Explosive Nucleosynthesis} 
\label{sec:explode} 

The previous section shows that stars with masses greater than the
Chandrasekhar mass continue to contract under the action of gravity
until their cores reach enormous temperatures and densities (Figure
\ref{fig:temptime}). The properties of these stellar cores
($T>10^{10}$ K and $\rho>10^{11}$ g cm$^{-3}$) are similar to the
conditions reached during Type Ia supernovae in our universe (e.g.,
see Figure 1 of Ref. \cite{bravo}) so that explosive nucleosynthesis
is expected to occur \cite{typeia,livarnett}.  However, the production
of any complex nuclei must proceed by first producing deuterium (or
another two-particle state) and then adding additional nucleons. 

In our universe, the temperatures reached in the stellar cores are
high enough that nuclear statistical equilibrium (NSE) can be realized
\cite{clifford,seitenzahl}.  In NSE, nuclear reactions are in detailed
balance, and the high densities help enforce this condition.  The mass
fractions of the various nuclear species are determined by minimizing
the Helmholtz free energy. As a result, high entropy environments
favor populations of lighter particles such as protons, neutrons, and
helium. Low entropy environments favor the production of larger
nuclei, those with the highest binding energy.

Although NSE provides a useful framework to estimate nuclear
abundances, nuclear reactions can take place out of equilibrium.
Moreover, NSE can only be reached if all of the relevant reactions
proceed fast enough. As shown in the following section, however, the
reaction rates for deuterium production are generally not fast enough
for the stellar core to reach NSE conditions (see Section
\ref{sec:nse}).  In this case, deuterium can still be produced, and
large nuclei are synthesized, but the abundances must be determined by
more detailed calculations (see Section \ref{sec:trinuke}), rather
than the standard statistical argument. In any case, the stellar core
will support a (small) standing population of deuterium, even though
the nuclei are unstable. These deuterium nuclei can interact further
to produce larger nuclei, which are stable, and thereby jump-start the
process of nucleosynthesis.

Although a detailed calculation of the nuclear yields for this
scenario is beyond the scope of this present paper, we can provide a
basic plausibility argument. The maximum temperature reached during
the collapse phase of massive stars is given by equation
(\ref{tmaxbig}). Significantly, this benchmark temperature
($\tmax\approx4.5$ MeV) is larger than three important energy scales:
[1] twice the electron rest mass (1.02 MeV), [2] the mass difference
between the proton and the neutron (1.29 MeV), and [3] the mass
difference between (unstable) deuterium nuclei and the corresponding
constituent particles (expected to be of $\sim1-2$ MeV). As the
temperature in the stellar core increases, pair production becomes
efficient, and the population of both neutrons and deuterons will
become significant due to considerations of nuclear statistical
equilibrium.  As a result, nuclear reactions will not necessarily be
suppressed due to the lack of a stable state for deuterium.  The
subsequent nucleosynthesis will then take place explosively as the
star continues to contract. In the end, heavy elements will be
produced and released into the background galaxy by the explosion.

For this scenario, it is important to note that explosive
nucleosynthesis can only operate over a limited range of parameter
space, where the masses and binding energies fall within a small
neighborhood of those in our universe.  The ordering of energy scales
outlined above requires that the mass difference between the proton
and neutron, and the nuclear binding energies, are of order 1 MeV. In
general, nuclear physics does not require that the binding energies of
nuclei (both positive and negative) must be comparable to the electron
mass or to the temperatures reached during stellar implosions. As a
result, universes with nuclear binding energies sufficiently different
from ours will not achieve explosive nucleosynthesis as outlined here.
In addition, although we consider unstable deuterium, we are also
assuming that nuclei with three nucleons ($A=3$), as well as $^4$He,
are bound.

When explosive nucleosynthesis is operative, the resulting nuclei can
be incorporated into subsequent generations of stars.\footnote{We  
are implicitly assuming that stars are forming and evolving inside a
galaxy, and that the explosion is contained within its gravitational
potential well.}  If these nuclei are primarily helium, then later
stellar generations can process them into carbon, which is useful both
for biology and for running the CNO cycle in those stars (see Section
\ref{sec:cno}).  If the explosion directly produces carbon and heavier
elements, then the metallicity of the galaxy will steadily increase,
and later generations of stars will also be able to operate via the
CNO cycle.

For completeness, we note that the temperature does not have to reach
the maximum value $\tmax$ in order for some nuclear processing to take
place.  In general, the abundance of deuterium is expected to be
suppressed by the multiplicative factor $\exp[-\Delta_d/T]$, where
$\Delta_d$ is a measure of the degree to which deuterium fails to be
bound (see Section \ref{sec:nse}). Since we are considering scenarios
where $\Delta_d\sim1$ MeV, as long as the temperature is not too far
below this value, some deuterium will be present and some nuclear
reactions will occur. However, once the density reaches the extremely
large values depicted in Figure \ref{fig:temptime}, the time scale for
further collapse becomes short.  In order to produce a standing
population of deuterium, weak interactions must be operative. As a
result, the production of deuterium can be suppressed due to the slow
rate of weak interactions. 


\section{Stellar Evolution through Triple-Nucleon Reactions} 
\label{sec:ppp} 

In the absence of stable deuterium, the synthesis of protons into
helium can take place through triple-nucleon reactions that are
roughly analogous to the triple-alpha process that produces carbon in
our universe \cite{clayton,kippenhahn}. For the triple-alpha process,
the transient population of unstable beryllium is determined by
Nuclear Statistical Equilibrium (NSE), so that we first consider the
limiting case where unstable deuterium arises in NSE (Section
\ref{sec:nse}). Because of the slow reaction rates and rapid decay of
deuterium, however, we find that NSE is generally not reached. As a
result, a more general treatment is developed by keeping track of the
population of free neutrons (Section \ref{sec:trinuke}). The resulting
reactions produce a transient population of deuterium that can
interact to produce $^3$He and eventually $^4$He. The evolution of
stars under the action of triple-nucleon processes is presented in
Section (\ref{sec:eresults}).

\subsection{Nuclear Statistical Equilibrium}  
\label{sec:nse}  

In this section, we consider the simplest case where a standing
population of deuterium is produced in NSE, and estimate the
properties required for the system of reactions to reach
equilibrium. At sufficiently high temperatures, the following
reactions take place: 
\be
p + p \to \, ^2\rm{He} + \gamma 
\label{ppdiproton} 
\ee
and 
\be
^2{\rm He} \to d + e^{+} + \nu_e \,,
\label{diprotondecay} 
\ee
where $^2$He is the diproton and $d$ is the deuterium nucleus. Since
the positron in this reaction will quickly find an electron and be
annihilated, we are left with a net reaction of the form
\be
e^{-} + p + p \to d + \gamma + \gamma + \nu_e \,. 
\label{net} 
\ee
Both the intermediate diproton nucleus ($^2$He) and deuterium ($^2$H)
are unstable under the assumptions considered in this paper.
However, these species will be present with some abundance due to
nuclear statistical equilibrium (NSE). We note that for analogous
reasons, evolved stars in our universe have a standing population of
$^8$Be, even though that nucleus is unstable. Most of the time, the
diproton and the deuteron will decay back to their constituent
particles.  Nonetheless, some deuterium will interact before decaying
via the reaction 
\be p + d \to \, ^3\rm{He} + \gamma \,.  \ee 
In our universe, this latter reaction produces an energy of 5.5 MeV:
the binding energy of helium-3 is $\sim7.7$ MeV, whereas the binding
energy of deuterium is $\sim2.2$ MeV.  Here we assume that $^3$He is
stable with a comparable binding energy, and that deuterium is
unstable.

In order to estimate the reaction rate for triple-nucleon processes, 
we must first find the expected abundance of unstable deuterium. 
In equilibrium, the chemical potentials must be equal on both 
sides of this equation so that 
\be
\mu_e + 2 \mu_p = \mu_d \,. 
\label{chemical} 
\ee
Note that the chemical potential is zero for the photon and is assumed
here to be zero for the neutrino.  We also use the usual condition for
kinetic equilibrium, which expresses the number density $n_X$ of
species $X$ in the form 
\be
n_X = g_X \left( {m_X T \over 2\pi} \right)^{3/2} 
\exp\left[ {\mu_X - m_X \over T} \right]\,,
\label{kinetic} 
\ee
where $g_X$ is the number of internal degrees of freedom. If we use
equation (\ref{kinetic}) to determine the chemical potentials of the
protons and deuterium in reaction (\ref{net}), the condition of
chemical equilibrium from equation (\ref{chemical}) takes the form
\be
n_d = n_p^2 \left({g_d \over g_p^2}\right) 2^{3/2} 
\left({2 \pi \over \mpro T}\right)^{3/2} 
\exp\left[{2 \mpro - m_d + \mu_e \over T} \right]\,.
\label{inter} 
\ee
To evaluate the chemical potential of the electron, we 
again invoke the condition of kinetic equilibrium (\ref{kinetic})
and the condition of charge neutrality, which implies that 
$n_e = n_p$. We obtain 
\be
\exp[ \mu_e/T ] = \exp[ m_e/T ] n_p g_e^{-1} 
\left({2\pi \over m_e T}\right)^{3/2}\,.
\ee
Using this result in equation (\ref{inter}), we find the 
following expression for the abundance of deuterium
\be
n_d = n_p^3 \left({g_d \over g_e g_p^2}\right) 2^{3/2} 
{(2 \pi)^3 \over (\mpro m_e)^{3/2} T^3} 
\exp\left[{2 \mpro + m_e - m_d \over T} \right]\,. 
\ee
Here we define 
\be
\Delta_d \equiv m_d - (2 \mpro + m_e) \,.
\label{deltadef} 
\ee
Since deuterium is unbound, $m_d > \mpro + m_n > 2\mpro + m_e$, 
so that $\Delta_d$ is a positive quantity. Although the quantity $\Delta_d$
could take a wide range of values, we are interested in the case 
where $\Delta_d={\cal O}$(1 MeV). In this regime, deuterium fails to 
be bound by an energy increment that is comparable to its actual 
binding energy in our universe. For much larger values of $\Delta_d$, 
we expect the binding energies of other relevant nuclei (e.g., 
helium-4) to change significantly. 

For convenience we define the scaled quantities 
\be
T_9 = {T \over 10^9 \,{\rm K}} 
\qquad {\rm and} \qquad 
n_{30} = {n_p \over 10^{30}\,{\rm cm}^{-3}} \,, 
\label{tnine} 
\ee
where the benchmark values are chosen to be comparable to the 
central temperatures and densities realized by stars undergoing
gravitational contraction (see Figure \ref{fig:temptime}). We 
also assume that the stars are pure hydrogen so that $n_p=n$. 
The abundance of deuterium in NSE is thus given by  
\be
\chi_d = {n_d \over n_p} = 2.3\times10^{-3} \, n_{30}^2 T_9^{-3} 
\exp\left[ -11.63 {\Delta_{\rm mev} \over T_9} 
\right]\,,
\label{nsedeut} 
\ee
where we have also defined $\Delta_{\rm mev}=\Delta_d$/(1 MeV). For 
example, if we fix $\Delta_{\rm mev}=1$ and $n_{30}=1$, then at high
temperatures $T_9=1$, we find a deuterium abundance of
$\chi_d\approx2\times10^{-8}$. In spite of the seemingly small
value, this deuterium abundance would lead to robust nuclear reactions.  
In our universe, the reaction rate for $p+d\to$ $^3$He is larger than
that of $p+p\to{d}$ by a factor of $\sim10^{18}$. As long as the
deuterium abundance is larger than about $10^{-18}$, nuclear reactions
can proceed fast enough to support the star.

The time required for the central core of a star to reach NSE
is determined by the slowest reaction rate. For the abundance 
of deuterium, the forward reaction is much slower than the 
decay of (unstable) deuterium back into its constituent parts. 
For the reaction of interest $pp\to{d}$, the cross section can 
be written in the form 
\be
\langle \sigma v \rangle_{pp} = 6.34 \times 10^{-39} 
\,\,{\rm cm}^3 \,\,{\rm sec}^{-1} \,\, T_9^{-2/3} f(T_9) 
\exp\left[-3.380 T_9^{-1/3} \right] \,,
\label{sigmavnse} 
\ee
where $T_9$ is defined by equation (\ref{tnine}) and 
where we have defined a function 
\be
f(T_9) = 1 + 0.123 T_9^{1/3} + 1.09 T_9^{2/3} + 0.938 T_9\,.
\ee

The reaction rate $\Gamma = n_p \langle \sigma v \rangle_{pp}$
and the corresponding time scale $\tau_{pp} = 1/\Gamma$ can be 
written in the form 
\be
\tau_{pp} \approx 1.6 \, {\rm yr} \,\,n_{30}^{-1} \,\,
\exp\left[3.380 T_9^{-1/3}\right] \,,
\ee
where we have used the benchmark value $T_9\approx1$ to evaluate the
polynomial part of the expression. Using this same value in the
exponential, we find a time scale $\tau_{pp}\approx47$ yr (for
$n_{30}=1$). This time scale is shorter than the time scales for
gravitational contraction ($\tau=1-10$ Myr), but much longer than 
the decay time for deuterium and the half-life of the neutron. As 
a result, NSE will not be maintained under most circumstances. 

We can also illustrate the difficulty in reaching NSE by considering
the equilibrium abundance of deuterium and comparing it to the NSE
value given by equation (\ref{nsedeut}). The net production rate for 
deuterium is given by 
\be
{d n_d \over dt} = {1\over2} n_p^2 \langle \sigma v \rangle_{pp} 
- \lambda_d n_d + \dots \,,
\label{drate} 
\ee
where $\lambda_d$ is the decay rate for deuterium. Here we assume that
the lifetime is comparable to that for $^8$Be in our universe, so that
$\lambda_d^{-1}\sim10^{-16}$ s. Equation (\ref{drate}) neglects
additional terms corresponding to the burning of deuterium into other
nuclei. These reactions take place on astrophysical time scales and
can be neglected for purposes of estimating equilibrium abundances.
In steady state, the time derivative $dn_d/dt=0$, so that the
equilibrium abundance of deuterium is given by 
\be
n_d = {1\over2} n_p^2 \langle \sigma v \rangle_{pp} \lambda_d^{-1} \,. 
\ee
Using the values $T_9=1$ and $n_{30}=1$, the density of deuterium 
becomes $n_d\approx 3 \times 10^4$, with a corresponding abundance 
$\chi_d=n_d/n_p\approx3\times10^{-26}$. Since this value is much 
smaller than the NSE value found in equation (\ref{nsedeut}), the 
system cannot maintain equilibrium. As a result, the abundance of 
deuterium must be determined out of equilibrium. This issue is addressed 
in the following section. 

\subsection{Triple-Nucleon Process including Free Neutrons} 
\label{sec:trinuke} 

In the previous section, we considered the abundance of unstable
deuterium in NSE and found that the required reaction rates are
generally too slow to reach equilibrium. This section considers an
alternate mechanism to bridge the $A=2$ mass gap in a universe without
stable deuterium. As outlined above, unstable deuterium will be
produced, even if the decay rate is too rapid for the system to reach
NSE.  These decays of deuterium follow the chain of reactions 
\be
\dstar \rightarrow p + n \rightarrow
p + p + e^- + \overline{\nu}_e,\label{rxn:path}
\ee
which occur quickly compared to evolutionary time scales of stars.
The second step in the path (\ref{rxn:path}) is free-neutron decay,
which has a mean lifetime $\tau_n=885.1\,{\rm s}$ in our universe
\cite{2012PhRvC..85f5503S,2013PhRvL.111v2501Y}.  The neutron lifetime
$\tau_n$ is indeed short compared to the lifetime of typical stars,
but is long compared to nuclear reaction time scales, implying that
the free neutrons from equation (\ref{rxn:path}) could participate in
nuclear reactions. In this section, we move beyond the assumption of
NSE and determine how stellar interiors evolve with a reservoir of
free neutrons.

In the presence of free neutrons, $^3$He can be synthesized from 
three free nucleons through a reaction of the form
\be
  n + p + p \rightarrow \,  ^3\rm{He} + \gamma.
\label{rxn:npp}
\ee
We call this process the triple-nucleon reaction. 
In analogy with the triple-alpha reaction \cite{1985A&A...149..239N},
we write the rate of production of $^3$He, $dY_3/dt|_{+}$, from
reaction (\ref{rxn:npp}) as 
\be\label{eq:dy3dt}
  \frac{dY_3}{dt}\biggr|_{+} = Y_nY_p^2\,\frac{\rho_b^2N_A^2}{2\Gamma(d)}
  \,\langle np\rangle\langle\dstar p\rangle,
\ee
where $Y_i$ is the abundance of species $i$, $\rho_b$ is the baryon
mass density, $\Gamma(d)$ is the radiative decay width of unstable
deuterium, and $\langle np\rangle$ and $\langle\dstar p\rangle$ are
the integrated products of cross section multiplied by speed
($\sigvbar$) for the reactions $n(p,\gamma)\dstar$ and
$\dstar(p,\gamma)\heiii$, respectively. Note that equation (\ref{eq:dy3dt}) 
is written in natural units. We will take the radiative
decay width $\Gamma(d)$ to be a parameter in our model. In our universe,
$\beryeight$ has a width $\Gamma(\beryeight)=6.8\,{\rm eV}$, and so we
will consider decay widths for deuterium with comparable values. For  
the two $\sigvbar$ quantities, we will take the known cross section of
$n(p,\gamma)d$ from Ref.\ \cite{kawano}, and that of
$d(p,\gamma)\heiii$ from the {\sl\small MESA} library.\footnote{For 
completeness, we note that there exists a reaction similar to that
of equation (\ref{eq:dy3dt}) in our universe, namely $d(p,n)2p$, which
has a known cross section $\langle\sigma v\rangle$ \cite{caughlanfowler}.  
The reverse channel of this reaction thus corresponds to the synthesis
of three individual nucleons into heavier bound states. However, this
work does not use the corresponding value of $\langle\sigma v\rangle$
for $2p(n,p)d$ from Ref.\ \cite{caughlanfowler} because the final
state does not contain $^3{\rm He}$ and because deuterium is unstable
in this current scenario.}   
Unlike the triple-alpha reaction, we will assume only thermal
non-resonant production of $^3$He through reaction (\ref{rxn:npp}).
With the absence of a resonance in the triple-nucleon reaction, the
`binding energy' of deuterium (denoted $B_d=m_d-m_p-m_n$) does not
enter into the expressions for $\langle np\rangle$ and $\langle\dstar
p\rangle$.  Furthermore, the expression for the production rate in
equation (\ref{eq:dy3dt}) does not contain an explicit dependence on
$B_d$.  There may exist an implicit dependence of the radiative decay
width on $B_d$, but we do not consider such models here.  Therefore,
our expression in equation (\ref{eq:dy3dt}) is independent of $B_d$.
In addition, the reverse photo-dissociation reaction, $\gamma + \heiii
\rightarrow p + p + n$, contains dependence on binding energies.
However, as the sum of the three individual nucleons is less massive
than a deuteron and a proton, there is no intermediate state with
deuterium.  As a result, only the binding energy of $^3$He
($B_3\simeq7.7\,{\rm MeV}$) is relevant in the reverse rate.

In order for the triple-nucleon reaction to operate in stars, some
pathway must be able to convert free protons into free neutrons. The
weak interaction offers a pathway through a modified form of the 
$p$-$p$ reaction 
\be
 p + p \rightarrow \dstar + e^+ + \overline{\nu}_e,
\ee
\be
\dstar \rightarrow p + n.
\ee
The above endothermic reaction chain converts two protons into one
proton and a neutron.  With the available neutron, the triple-nucleon
reaction from equation (\ref{rxn:npp}) can take place in stars.  Given
the stellar time scales and the languidness of the triple-nucleon
reaction, we also must allow for the neutron to decay into a proton.
We modify the nuclear network of the stellar code {\sl\small MESA} in
regard to the following reactions 
\be
p + p \rightarrow p + n + e^+ + \nu_e,\label{rxn:3nuc_1}
\ee
\be
n \rightarrow p + e^- + \overline{\nu}_e,\label{rxn:3nuc_2}
\ee
\be
n + p + p \leftrightarrow\, ^3{\rm He} + \gamma.\label{rxn:3nuc_3}
\ee
For reaction (\ref{rxn:3nuc_1}), we use the identical $\sigvbar$ in our
universe for the reaction $p(p,e^+\nu_e)d$, and demand that the final
state is a neutron and proton instead of a deuteron.  For neutron
decay, we simply set the rate equal to $1/\tau_n$.  The triple-nucleon
reaction uses the expression in equation (\ref{eq:dy3dt}) for $\sigvbar$.

\subsection{Stars Powered by the Triple-Nucleon Process} 
\label{sec:eresults} 

\begin{figure}
\begin{center}
\includegraphics[scale=0.67]{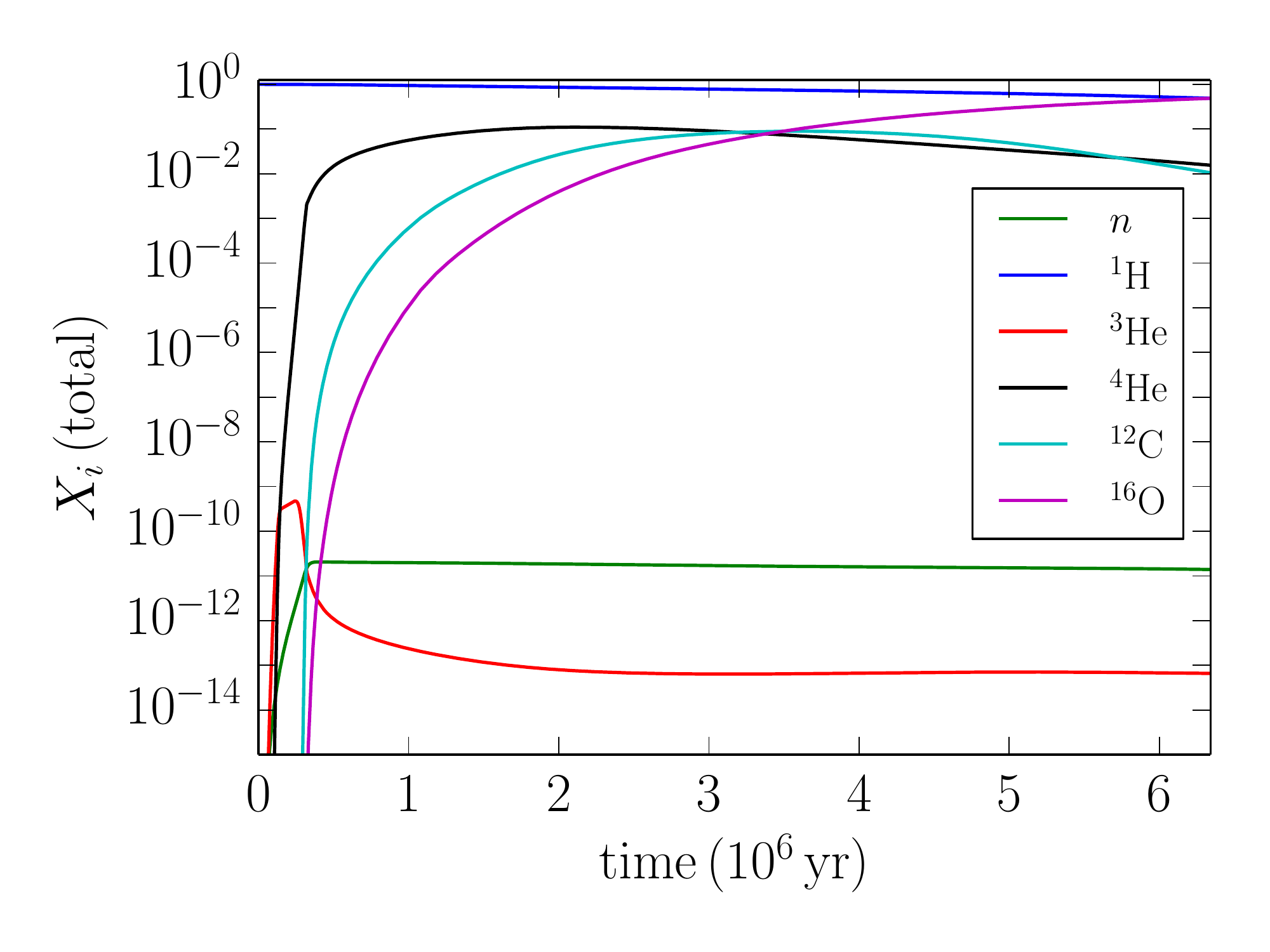}
\end{center}
\caption{\label{fig:abunds_vs_time} Mass fractions versus time for a    
15 $M_\odot$ star in a universe without stable deuterium. The initial
metallicity of the star is zero. The model employed here uses the set
of reactions (\ref{rxn:3nuc_1}) -- (\ref{rxn:3nuc_3})  
with $\Gamma(d)=10.0\,{\rm eV}$ to synthesize $^3$He.  }
\end{figure}

To illustrate the evolution of stars operating via the triple-nucleon
reaction, we first present the results from the simulation of a
$M_\ast$ = 15 $M_\odot$ star with zero initial metallicity.  The
radiative decay width of deuterium is set to $\Gamma(d)=10.0$ eV. 
All of these results were obtained using modified versions of the
{\sl\small MESA} computational package \cite{paxton,paxtonb}. For
these simulations, the $p$-$p$ chain is replaced by the reactions
described in the previous section and the CNO cycle is inoperative,
but the higher order (e.g., helium burning) reactions are the same as
those in our universe.  Figure \ref{fig:abunds_vs_time} shows the
evolution of the mass fractions for a variety of nuclei as a function
of time, where the values are averaged over the total volume of the
star.  Similar to stars in our universe, the abundances of $^4$He,
$\carbonxii$, and $\oxvi$ rise and fall with increasing time, as
different nuclear reactions become important. In addition, there is a
standing population of $^3$He which is continuously incorporated into
$^4$He, and replenished by the triple-nucleon reaction.

One interesting feature of the evolution shown in
Figure\ \ref{fig:abunds_vs_time} is the presence of a sea of neutrons,
which have an mass fraction of order $X_n\sim10^{-11}$, averaged over
the total volume of the star.  In this model, the transmutation of a
free neutron only has two pathways: incorporation into a $^3$He
nucleus via the triple-nucleon reaction or beta-decay into a free
proton.  Once heavier elements are present (e.g., $\carbonxii$ or
$\oxvi$), free neutrons could capture on the larger $Z$ nuclei.  The
result of this process would be to slow the synthesis of $^3$He.
However, we did not include any neutron capture reactions in the model
used in Figure \ref{fig:abunds_vs_time}.  In addition, we did not
include a two-neutron variant of the triple-nucleon reaction within
{\sl\small MESA}.  The neutron abundance is so small that the
contribution of $2n(p,\gamma)\hiii$ is insignificant. Similarly, we 
did not include a three-proton variant of the triple-nucleon reaction.
That reaction would require the weak interaction and would be slower
than the $npp$ reaction once the free neutrons have a standing
population.

\begin{figure}
\begin{center}
\includegraphics[scale=0.67]{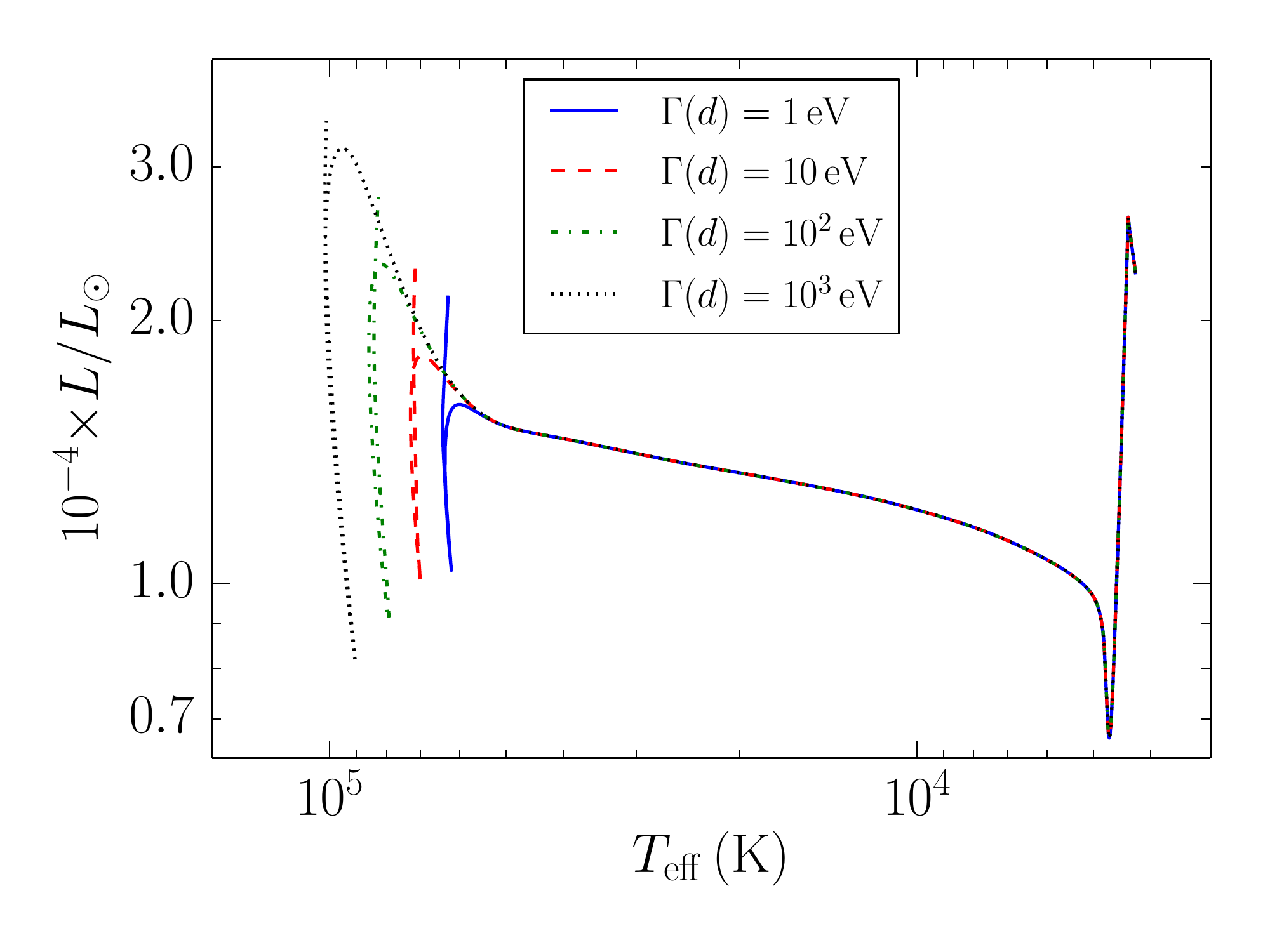}
\end{center}
\caption{\label{fig:NSE_vs_3nuc} Evolutionary tracks in the H-R 
diagram for a 15 $M_\odot$ star in a universe without stable deuterium.
The tracks show the results for different values of the parameters that
specify the triple-nucleon reactions. The four curves follow the 
evolution in the diagram for the triple-nucleon model of Section 
\ref{sec:trinuke} for various values of the width $\Gamma(d)$.  }
\end{figure}

Next we consider the evolution of stars in the H-R diagram operating
through the triple-nucleon reactions from Section \ref{sec:trinuke}.
Figure\ \ref{fig:NSE_vs_3nuc} shows a comparison of the tracks in the
H-R diagram for stellar models with mass $M_\ast=15M_\odot$. The solid
blue, dashed red, dash-dot green, and dotted black curves depict the
evolution of stellar models using the triple-nucleon reaction with
varying values of the decay width $\Gamma(d)=1-10^3$ eV.  We terminate
the evolutionary tracks after the mass fraction of metals in the
stellar core reaches 50\% (where metals are defined to be all nuclei
with $A>5$).

The stellar models illustrated in Figure \ref{fig:NSE_vs_3nuc} show
broadly similar evolution with some interesting differences. All of
the models follow essentially the same tracks in the H-R diagram for
the beginning phases of evolution. The tracks continue until the
central core of the star is hot enough for enough hydrogen fusion to
occur rapidly enough --- through the triple-nucleon process --- to
support the star against further contraction. The surface temperature
at the onset of nuclear burning depends on the decay width
$\Gamma(d)$. Larger values of the width $\Gamma(d)$ correspond to
faster decay of deuterium, which in turn require higher central
temperatures for sustained nuclear reactions and produce correspondingly
higher surface temperatures.  The luminosity of the star also tends to
increase with increasing values of the decay width $\Gamma(d)$,
although the variation is relatively small (a factor of $\sim2$) and
the tracks move up and down in luminosity with time. Notice also that
a 15 $M_\odot$ star in our universe has luminosity
$L_\ast\sim10^4L_\odot$, roughly comparable to the luminosities
displayed in Figure \ref{fig:NSE_vs_3nuc}.

\section{Stellar Evolution including the CNO Cycle} 
\label{sec:cno} 

Another way for stars to generate energy in the absence of stable
deuterium is through the CNO cycle. This chain of reactions requires
somewhat higher temperature and density than the $p$-$p$ reaction
chain, but it does not require a stable deuterium nucleus as an
intermediate state. On the other hand, the cycle only operates with a
minimum abundance of the element carbon, where the requisite values
are determined below. As a result, some carbon nuclei must be produced
by alternative means, either through explosive nucleosynthesis during
the collapse of stellar cores or through the triple-nucleon process
(and subsequent triple-alpha process).  As we show here, however,
the required carbon abundance can be much smaller than the typical
values in our universe.

In this section, we first present our treatment of the CNO cycle
(Section \ref{sec:cnoreactions}) and then show how stars can evolve
with no deuterium and trace amounts of carbon via the CNO cycle
(Section \ref{sec:cnoresults}). We then consider how stars evolve
during the action of both the triple-nucleon process and the CNO cycle
(Section \ref{sec:combine}) and finally present a comparison of the
different scenarios for hydrogen burning considered in this paper
(Section \ref{sec:comparison}).

\subsection{The CNO Reaction Chain} 
\label{sec:cnoreactions} 

With non-zero abundances of carbon, the CNO cycle proceeds through 
the following set of nuclear reactions: 
\be
^{12}{\rm C}+p \to \, ^{13}{\rm N}+\gamma
\label{cnofirst} 
\ee
\be
^{13}{\rm N} \to \, ^{13}{\rm C} + e^+ + \nu_e
\label{betaone} 
\ee
\be
^{13}{\rm C}+p \to \, ^{14}{\rm N}+\gamma
\ee
\be
^{14}{\rm N}+p \to \, ^{15}{\rm O}+\gamma
\label{slowest} 
\ee
\be
^{15}{\rm O} \to \, ^{15}{\rm N} + e^+ + \nu_e 
\label{betatwo} 
\ee
\be
^{15}{\rm N}+p \to \, ^{12}{\rm C}+ \, ^4{\rm He} \,.
\label{cnolast} 
\ee

Note that the reactions (\ref{betaone}) and (\ref{betatwo}) represent
beta decay. Use of this particular set of reactions implicitly assumes
that the beta decays have time to occur before the parent nuclei
interact further.  This approximation is valid for stellar
nucleosynthesis in the solar core.  In the present context, the
reaction rates are likely to be smaller (than in our universe) due to
the reduced abundances of the CNO nuclei. In the event that the star
in question has enough of these elements, and high enough temperature
and density, so that the CNO reactions proceed rapidly, the star can
generate energy through additional sets of nuclear reactions.  For
this present treatment, we focus on the simpler case corresponding 
(only) to reactions (\ref{cnofirst}) to (\ref{cnolast}). 

The slowest reaction in the CNO cycle is given by equation
(\ref{slowest}) for $^{14}$N + $p\to$ $^{15}$O. This reaction thus
determines the overall rate at which energy is generated. The net 
energy generation rate for the entire cycle can then be written in 
the form 
\be
\epsilon_{\rm{CNO}} \approx 10^{26} 
\,{\rm erg}\,\,{\rm g}^{-1}\,{\rm sec}^{-1} \rho X_H X_C T_9^{-2/3} 
\exp\left[ - {15.23 \over T_9^{1/3}}\right] \,,
\label{epsiloncno} 
\ee
where $X_C$ is the mass fraction of carbon. For comparison, the energy 
generation rate for the $p$-$p$ chain can be written in the form 
\be
\epsilon_{pp} \approx 10^{4} 
\,{\rm erg}\,\,{\rm g}^{-1}\,{\rm sec}^{-1} \rho X_H^2 T_9^{-2/3} 
\exp\left[ - {3.381 \over T_9^{1/3}}\right] \,. 
\label{epsilonpp} 
\ee
As a point of comparison, we can equate these two expressions and 
find the conditions required for the CNO cycle to provide as much 
energy the $p$-$p$ chain does for stars in our universe. We thus find 
\be
10^{22} X_C = X_H \, \exp\left[ {11.85 \over T_9^{1/3}}\right] \,. 
\label{carbonconstraint} 
\ee

The largest temperature that can be achieved in a star --- in the
absence of collapse --- is given by equation (\ref{tmaxdegen}) for a
given stellar mass. This equation assumes that the stars are below the
Chandrasekhar mass limit ($M_\ast/M_\odot<5.6$), but it provides a
good benchmark for all stellar masses. Using this result, and taking 
$X_H=1$, the required carbon mass fraction can be written in the form  
\be
\log_{10} X_C \approx -22 + 12.7 (M_\ast/M_\odot)^{-4/9} \,.
\label{mincarbon}
\ee
For the largest star that can be supported by degeneracy pressure, and
hence does not collapse, the mass $M_\ast/M_\odot\approx5.6$ and the
minimum required carbon abundance $X_C \approx 10^{-16}$. For a solar
mass star, the minimum carbon abundance is larger, 
$X_C \approx 5 \times 10^{-10}$. Note that these abundances are much
smaller than those corresponding to Solar metallicity. This difference
arises because the stars in question can achieve maximum temperatures
$T\sim7\times10^8$ K, which are much larger than the operating
temperature of solar-type stars.

\begin{figure}
\begin{center}
\includegraphics[scale=0.67]{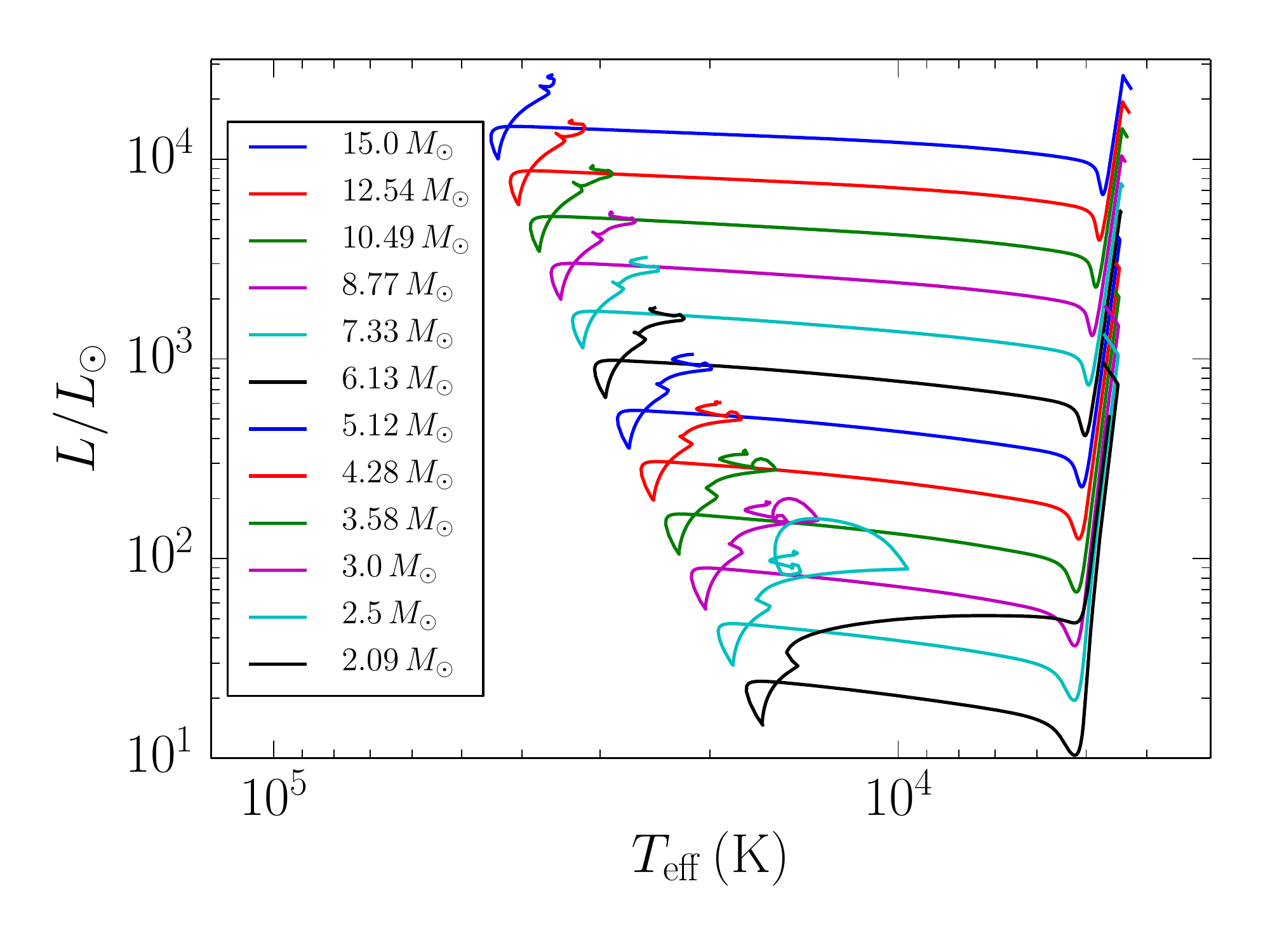}
\end{center}
\caption{H-R diagram for stars evolving under the action of the CNO 
cycle with metallicity $Z=10^{-8}$. In this scenario, both the $p$-$p$
chain and the triple-nucleon process are assumed to be inoperative. 
The ends of the evolutionary tracks correspond to configurations 
where the mass fraction of elements with $A>5$ exceeds 50\% in the 
stellar core. }
\label{fig:hrcno} 
\end{figure}

\subsection{Stars Powered by the CNO Cycle} 
\label{sec:cnoresults} 

To illustrate the effectiveness of the CNO cycle, we have run the
{\sl\small MESA} stellar evolution code for the following scenario.
Deuterium is assumed to be unstable. As a result, the standard $p$-$p$
chain of reactions (which powers low-mass stars in our universe) is
assumed to be inoperative. In order to isolate the viability of the
CNO cycle, we also assume that the triple-nucleon reaction (see Section
\ref{sec:ppp}) is not working. The metallicity of the star is
taken to be small, $Z=10^{-8}$. By definition, the value of $Z$
determines the mass fraction of all elements heavier than helium. For 
the sake of definiteness, we assume that the relative abundances of
these elements are the same as the cosmic abundances in our universe, 
although the mass fraction of carbon is the crucial quantity.  In
addition, we set the initial mass fraction $Y$ of helium to be zero,
so that we are implicitly assuming that Big Bang
Nucleosynthesis is also ineffective. This low but nonzero value of $Z$
could be produced, for example, by an early generation of massive
stars that evolve through gravitational contraction (Section
\ref{sec:gravity}) and then explode as supernovae when they enter into
their collapse phase after a few Myr (Figure \ref{fig:temptime}). A
metallicity $Z=10^{-8}$ requires about 100 solar masses of carbon for
a galaxy with mass comparable to the Milky Way.

Under the conditions outlined above, the evolution of the stars in the
H-R diagram is shown in Figure \ref{fig:hrcno}. The evolution is much
like those of stars in our universe. The stars begin with fully
convective interiors, so that they first contract on nearly vertical
tracks in the H-R diagram. After becoming radiative, the stars then
evolve on nearly horizontal tracks; they move to the left in the H-R
diagram, with increasing surface temperatures, until hydrogen burning
commences through the CNO cycle. The onset of nucleosynthesis produces
a well-defined main sequence over the range of stellar masses, as
delineated by the left boundary of the envelope of tracks shown in
Figure \ref{fig:hrcno}. Note that the formation timescale for stars is
of order 0.1 Myr. For high mass stars, with $M_\ast$ $>7M_\odot$, the
time required for the stars to evolve to a hydrogen burning
configuration is shorter than the formation time, so that these stars
will first become optically visible with their main-sequence properties. 

\begin{figure}
\begin{center}
\includegraphics[scale=0.67]{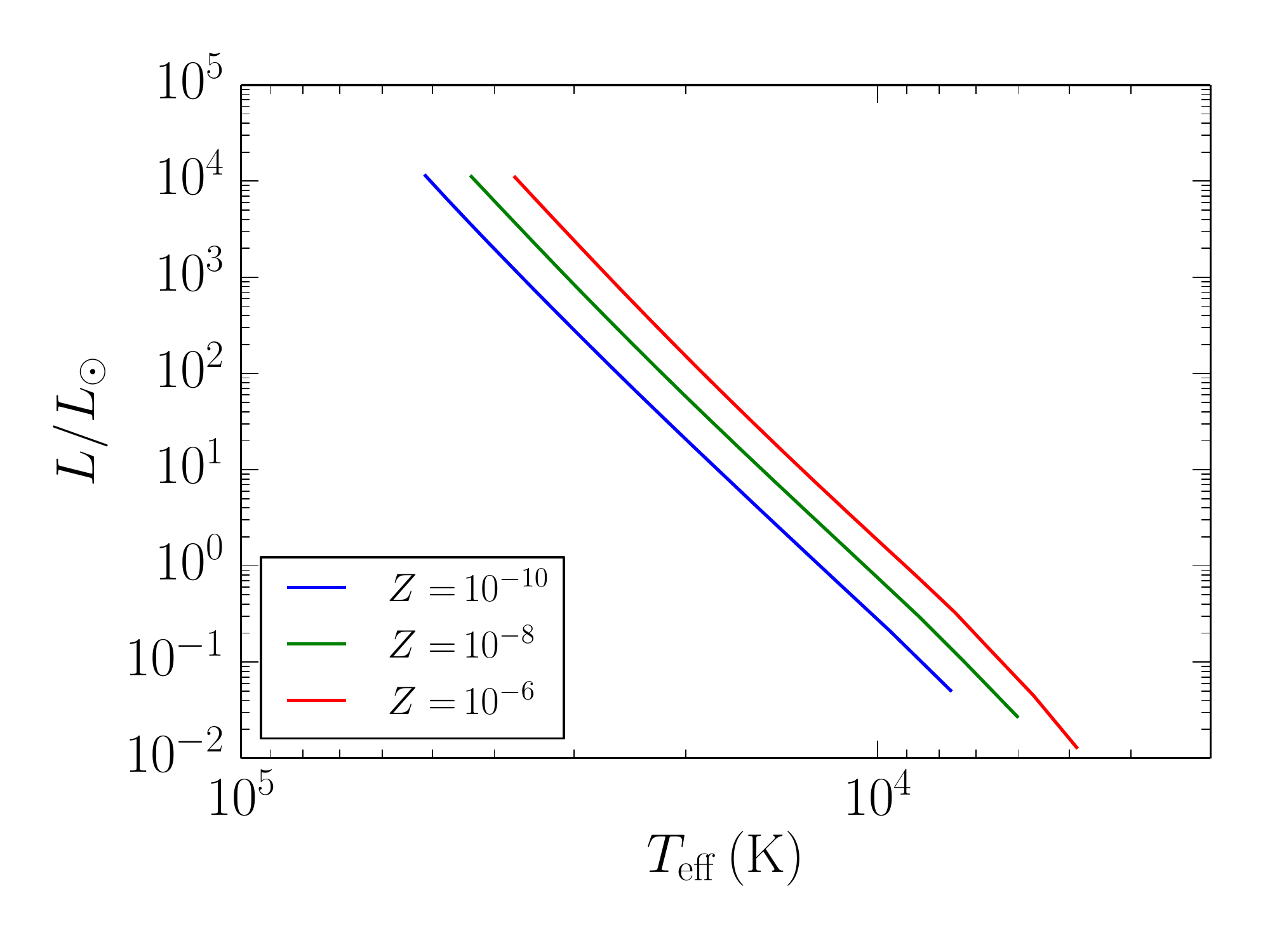}
\end{center}
\caption{H-R diagram for stars of varying metallicity evolving under  
the action of the CNO cycle only. The curves show the Zero-Age
Main-Sequence (ZAMS) for stars with metallicity $Z=10^{-6}$ (red),
$10^{-8}$ (green), and $10^{-10}$ (blue). For this scenario, both 
the $p$-$p$ chain and the triple-nucleon process are assumed to be
inoperative. }
\label{fig:cnozams} 
\end{figure}

The resulting main-sequence for these CNO stars is characterized by
somewhat higher luminosity and significantly higher surfaces
temperatures compared to stars in our universe. The low metallicity
requires the stars to contract to higher central temperatures for the
CNO cycle to operate (see equation [\ref{carbonconstraint}]) and also
leads to lower overall opacity. This latter property allows
radiation to escape from the star more readily and requires the stars
to be hotter and brighter, for a given stellar mass, than stars with
ordinary metallicity (e.g., $Z \sim Z_\odot \sim 0.02$). To explore
this issue further Figure \ref{fig:cnozams} shows the zero-age
main-sequence for stars with varying metallicity in the range
$Z=10^{-6}-10^{-10}$. As the value of $Z$ decreases, the main-sequence
falls farther to the left in the H-R diagram. This trend reflects the
higher surface temperatures for stars with little metals, as indicated 
by equation (\ref{mincarbon}). 

These CNO-only stars burn hotter than those in our universe and thus
have a correspondingly shorter main-sequence lifetime (although the
lifetime is still longer than that for stars with no nuclear
reactions, as shown in Figure \ref{fig:habtime}). The low values of
metallicity considered here apply only to the first generation of
stars. After finishing their hydrogen burning phase (via the CNO
cycle), sufficiently massive stars will continue to evolve and produce
ever-heavier elements, analogous to how stellar evolution takes place
in our universe. Subsequent generations of stars will thus have higher
metallicity and will appear quite similar to those in our universe.
The key issue is not the stability of deuterium, but rather the
existence of bound states for a suite of heavier elements. The element
carbon is especially important, as it is usually considered as the
basis for life and, in this context, plays a vital role in catalyzing
nuclear reactions in stars.

\subsection{Combined Triple-Nucleon Process and CNO Cycle} 
\label{sec:combine} 

\begin{figure}
  \begin{center}
    \includegraphics[scale=0.67]{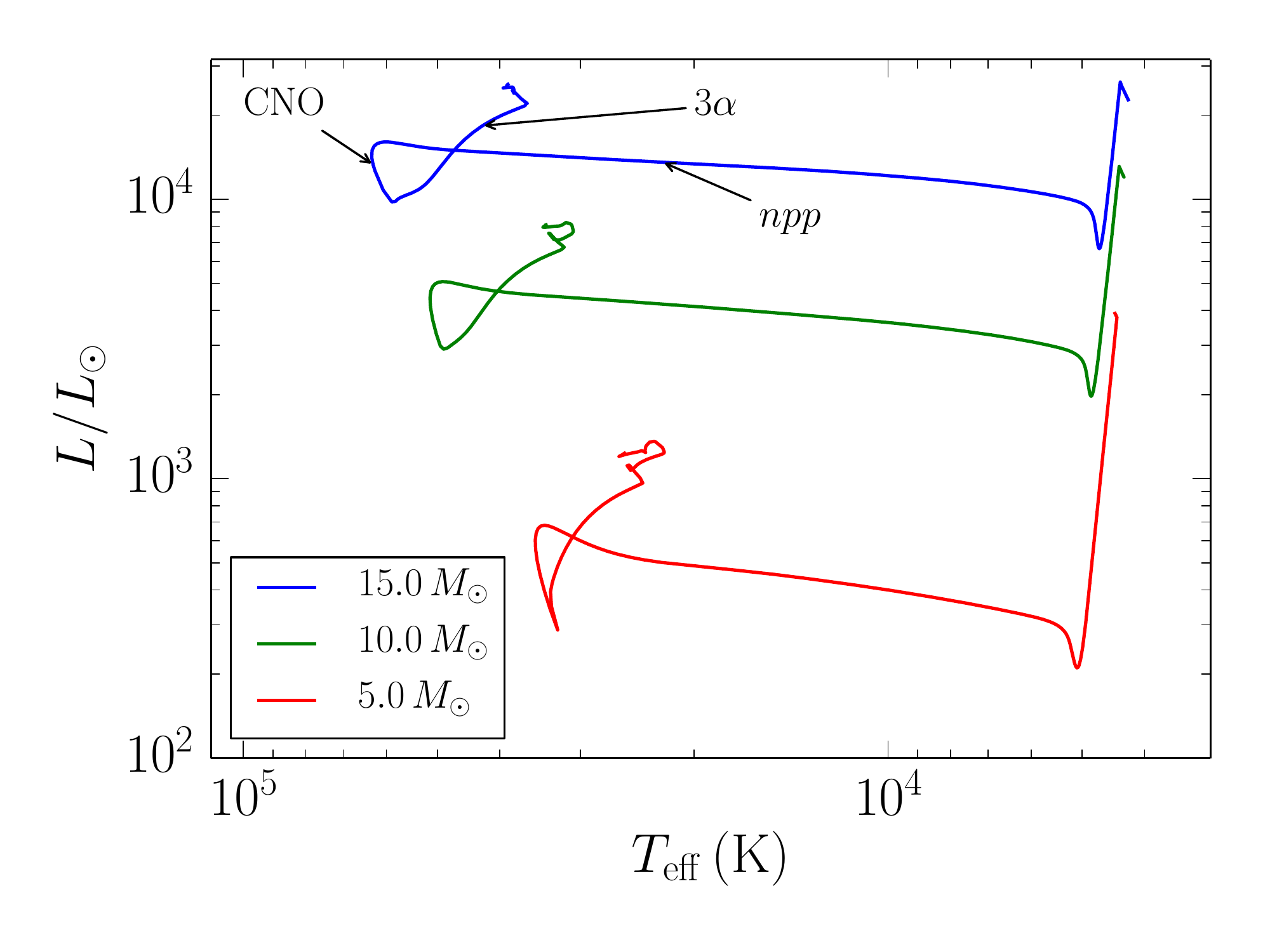}
  \end{center}
\caption{\label{fig:3nuc_set} Tracks in the H-R diagram for stars 
with three different masses in a universe without stable deuterium.
Initially, the stars have zero metallicity. The model employed here
uses both the triple-nucleon reaction and CNO reactions.  Labels on
the 15 $M_\odot$ curve correspond to different burning regimes,
namely: $npp$ when the core $\hydroi$ fraction falls to $99\%$; CNO
when the reaction luminosity for CNO surpasses that of $npp$; and
$3\alpha$ when the core $\carbonxii$ mass fraction rises to $1\%$.  }
\end{figure}

We now consider the evolution of stars where both the triple-nucleon
process and the CNO cycle are operational.  Figure \ref{fig:3nuc_set}
shows the resulting tracks in the H-R diagram for three stars of
different masses, each starting at zero metallicity.  We employ the
set of reactions (\ref{rxn:3nuc_1}) -- (\ref{rxn:3nuc_3}) with
$\Gamma(d)=10.0\,{\rm eV}$ to initially synthesize $^3$He (and
subsequently $^4$He) from free protons.  We also include the reactions
from the CNO cycle (see Section \ref{sec:cnoreactions}) in this model.
The labels on the H-R track for the $15\,M_\odot$ star correspond to
various burning regimes.  First, the triple-nucleon reaction burns
single nucleons into $^3$He.  At the point labeled $npp$, the star has
burned $1\%$ of the $\hydroi$ in its core into $^3$He.  The $^3$He is
incorporated into $^4$He, which is burned into $\carbonxii$.  Once
there is a substantial population of $\carbonxii$ available, the CNO
cycle more efficiently burns $\hydroi$ into $^4$He.  At the point
labeled CNO, the reaction luminosity (power output for a given set of
reactions integrated over the total volume of the star) for CNO
surpasses that of the triple-nucleon reaction.  Towards the end of the
track, we label the point $3\alpha$ when the mass fraction of
$\carbonxii$ accounts for more than $1\%$ of the core.  $\carbonxii$
is required for the CNO cycle to operate.  As we started with zero
metallicity, the triple-alpha reaction must produce $\carbonxii$
before the CNO cycle commences.  We labeled $3\alpha$ after CNO to
show the regimes where each reaction or set of reactions dominates,
but the $3\alpha$ reaction is actually in operation before the labeled
point in the HR track.  The CNO cycle requires very little
$\carbonxii$ to operate.  Indeed, at the point labeled CNO, the mass
fraction of $\carbonxii$ in the core is $X_{12}\simeq3\times10^{-14}$.

\subsection{Comparison of the Nuclear Models} 
\label{sec:comparison} 

\begin{figure}
\begin{center}
\includegraphics[scale=0.67]{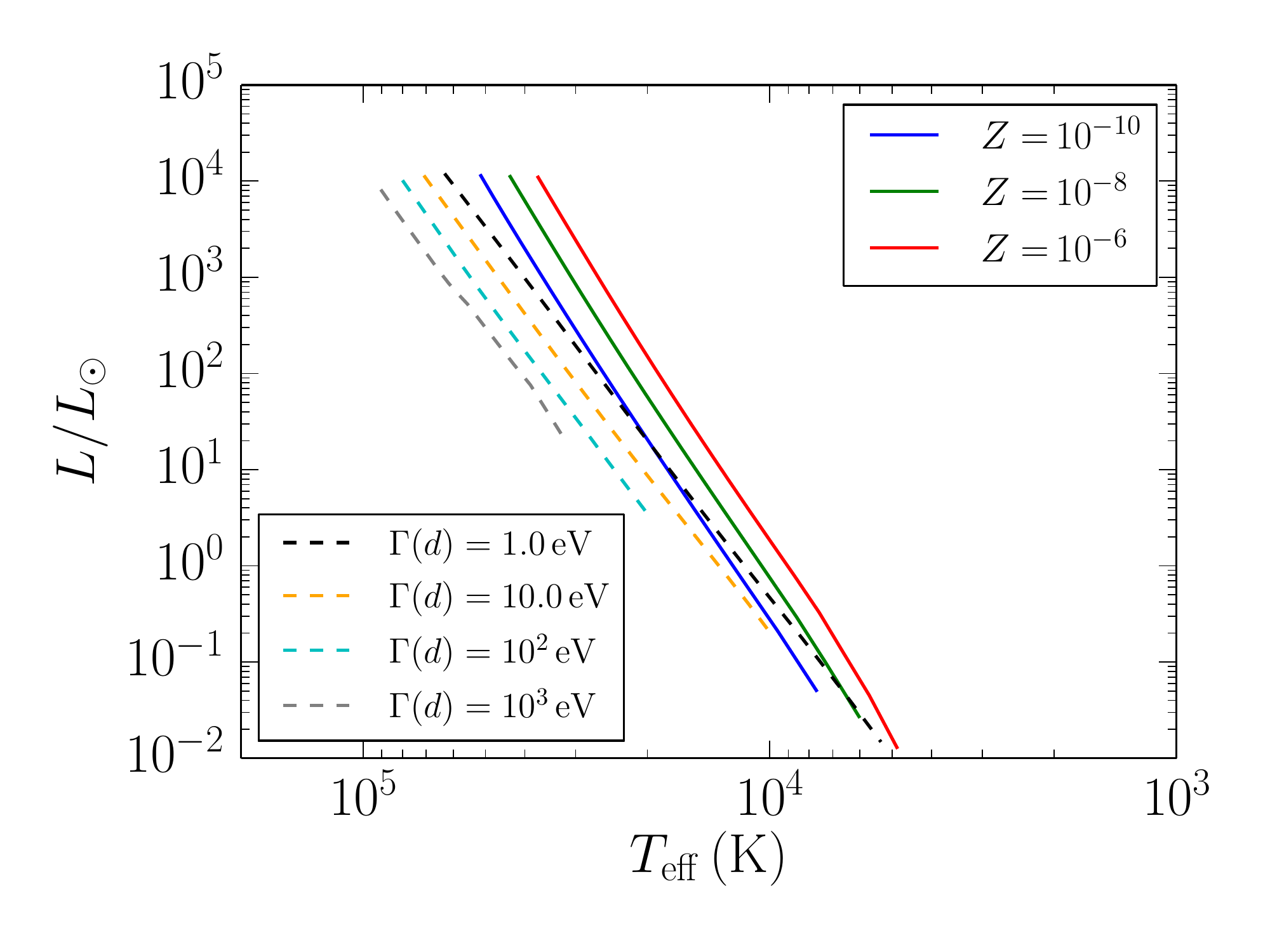}
\end{center}
\caption{\label{fig:ZAMS} Zero-Age Main-Sequence for the different 
stellar models in a universe without stable deuterium.  The masses
range from $15\,M_\odot$ down to $0.5\,M_\odot$, depending on the
models and parameters.  The solid curves correspond to stars that 
only burn $\hydroi$ using the CNO cycle for metallicities $Z$ = 
$10^{-6}$ (red), $10^{-8}$ (green), $10^{-10}$ (blue). The dashed 
curves correspond to stars that employ the triple-nucleon reaction
for widths $\Gamma(d)$ = $1-10^3$ (from right to left).  }
\end{figure}

Figure \ref{fig:ZAMS} shows the Zero-Age Main-Sequences (ZAMS) for the
different nuclear models considered in this work. For the sake
of definiteness, we define the ZAMS to be when $5\%$ of the initial
$\hydroi$ has been synthesized into heavier elements.  The mass range
depends on the specific model and parameter value.  CNO models only
employ the CNO cycle to synthesize $^4$He from free protons.  Those
models depend on the initial metallicity parameter, $Z$, and are shown
in Figure\ \ref{fig:ZAMS} as solid lines.  Decreasing the metallicity
slows the CNO reactions and pushes the ZAMS to higher effective
temperatures.  In addition, the stellar mass required for CNO burning
increases with decreasing $Z$.  For $Z=10^{-6}$, $0.5\,M_\odot$ stars
are able to burn $\hydroi$.  For $Z=10^{-10}$, stars must have larger
masses, upwards of $0.72\,M_\odot$.  As a result, the curve marking
the ZAMS for low metallicity $Z$ is not as long as that for higher $Z$
in Figure\ \ref{fig:ZAMS}.  The CNO models do not require as high of
temperatures as the triple-nucleon models, at least for large masses.
Those models show that increasing the radiative decay width,
$\Gamma(d)$, raises the temperatures needed for $\hydroi$ burning.
Similar to the trends for the CNO models, increasing $\Gamma(d)$
raises the minimum stellar mass required for $\hydroi$ burning.  The
lowest mass plotted for the $\Gamma(d)=1.0\,{\rm eV}$ model is
$M_\ast=0.72\,M_\odot$, as opposed to $1.0\,M_\odot$ for
$\Gamma(d)=10^3\,{\rm eV}$.  Notice that for small masses, the
triple-nucleon reaction with $\Gamma(d)=1.0\,{\rm eV}$ is competitive
with the CNO cycle. 

\begin{figure}
\begin{center}
\includegraphics[scale=0.67]{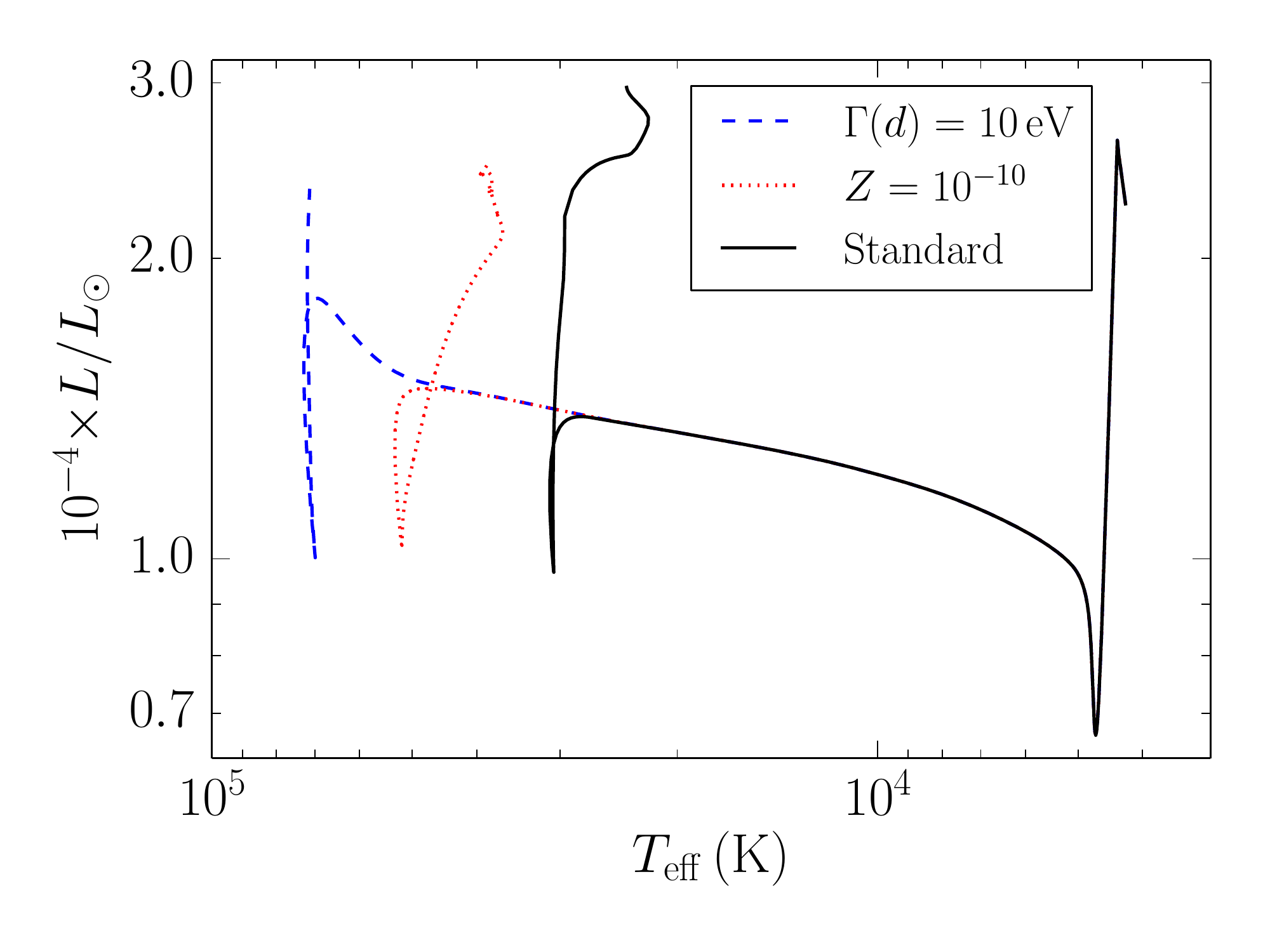}
\end{center}
\caption{\label{fig:trackcompare} Evolutionary tracks in the H-R
diagram for a star with mass $M_\ast=15M_\odot$ operating with 
three different nuclear reaction chains for hydrogen burning.
The three cases shown include the standard reaction chains in 
our universe (solid black curve), the CNO cycle with metallicity 
$Z=10^{-10}$ (dotted red curve), and the triple-nucleon reaction 
with decay width $\Gamma(d)$ = 10 eV (dashed blue curve). } 
\end{figure}

In general, the slopes of the ZAMS are roughly parallel to each other
for differing parameter values in a given model. Decreasing the
metallicity $Z$ in the CNO models, or increasing the decay width
$\Gamma(d)$ in the triple-nucleon models, respectively, moves the ZAMS
towards higher effective temperatures, while maintaining the same
slope. The slope of the ZAMS provides a measure of the effectiveness
of the nuclear reactions in supporting the star.  Figure
\ref{fig:ZAMS} shows that the main-sequences for the CNO models have
somewhat larger slopes than those of the triple-nucleon models. More
specifically, the slope of the ZAMS determines the degree to which
increasing the stellar mass increases the power output of the
star. Because the slope is larger for the CNO models, the reaction
rates increase more rapidly with stellar mass (compared to the
triple-nucleon models). In addition, as the reactions become less
effective, either through decreasing metallicity or increasing decay
width, the main-sequences shift to the left in the H-R diagram. The
result is higher surface temperatures for the same luminosity.

We can also compare how the different nuclear processes considered in
this paper affect the evolutionary tracks in the H-R diagram for
individual stars. Figure \ref{fig:trackcompare} show the resulting
tracks for stars with mass $M_\ast=15M_\odot$ and three types of
nuclear reaction chains.  The black solid curve shows the evolutionary
track for a star operating with all of the ordinary nuclear reactions
in our universe. The dotted green curve shows the track for a star
where hydrogen burning occurs only through the CNO cycle and where the
metallicity is low ($Z=10^{-10}$).  Finally, the dashed blue curves
shows the evolutionary track for a star burning hydrogen through the
triple-nucleon process from Section \ref{sec:trinuke}, where the
radiative decay width of deuterium is taken to be $\Gamma(d)=10$
eV. For all of these cases, the higher order nuclear reactions (e.g.,
the triple-alpha process for helium burning) are assumed to be the
same as in our universe. This assumption can be relaxed in future
work, but the number of scenarios is large (and beyond the scope of
this paper).

Along the sequence of models shown in Figure \ref{fig:trackcompare},
from the ordinary reactions of our universe to triple-nucleon
reactions with a large decay width, hydrogen burning becomes
increasingly difficult. In all cases, however, the nuclear reaction
rate must be large enough to provide pressure support for the entire
stellar mass.  As a result, the star must contract further to provide
increasingly higher temperatures and densities in its core.  The stars
thus follow their radiative pre-main-sequence-like tracks farther to
the left in the H-R diagram before adjusting downward to lower
luminosities. This behavior leads the stars to become hotter in
effective temperature as well as in their cores. The stars also become
somewhat brighter along this sequence of increasingly difficult
hydrogen burning, but only by a factor of $\sim2$. The differences in
effective temperature are more pronounced. 

\begin{figure}
\begin{center}
\includegraphics[scale=0.67]{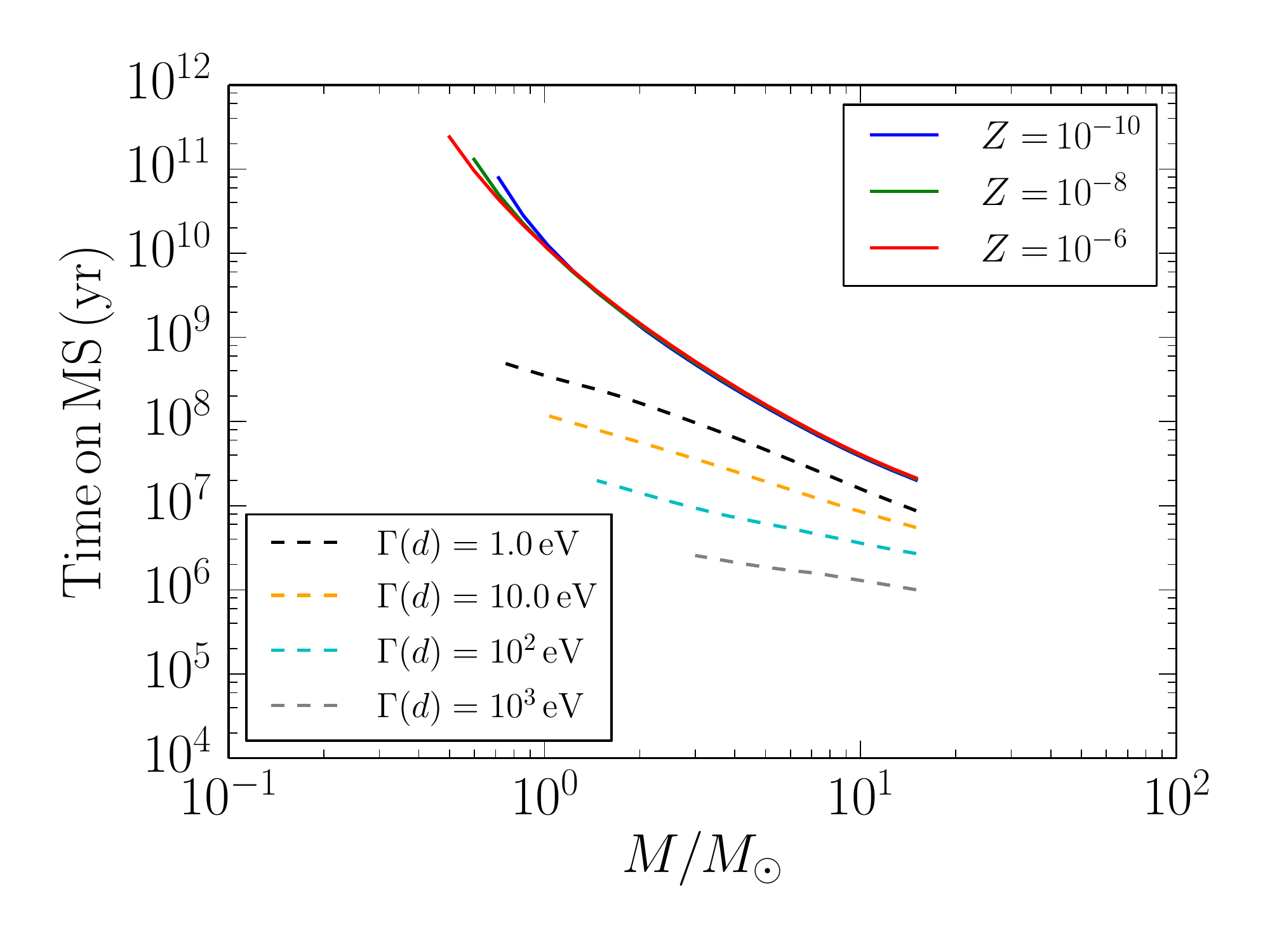}
\end{center}
\caption{\label{fig:timecompare} Main sequence lifetimes for stars
with varying mass operating with the triple nucleon process (dashed
curves) and with the CNO cycle (solid curves). The dashed curves for
the triple nucleon process correspond to stars with decay widths
$\Gamma(d)$ = 1 eV (black), 10 eV (gold), 100 eV (cyan), and $10^3$ eV
(gray).  The solid curves for the CNO cycle correspond to initial
metallicities $Z$ = $10^{-6}$ (red), $10^{-8}$ (green), and $10^{-10}$
(blue). }
\end{figure}

Figure \ref{fig:timecompare} shows the main sequence (hydrogen
burning) lifetimes for stars as a function of mass for the two types
of nuclear processes considered in this work.  For the sake of
definiteness, the starting condition is defined as the time when the
star burns 5\% of the hydrogen (by mass) in its core, where the core
is defined to be the inner 10\% of the star (again by mass). The
ending condition is defined as the time when the star burns 95\% of
the hydrogen (by mass) in its core, or when the core reaches a point
where more than 50\% of the mass is sequestered in nuclei with $A>5$.
Note that the minimum mass for hydrogen burning varies with the
nuclear process under consideration, so that not all masses are
represented. The solid curves show the main sequence lifetimes for
stars operating through the CNO cycle for metallicities in the range
$Z=10^{-6}-10^{-10}$. The dashed curves show the lifetimes for stars
operating through the triple nucleon process, where the decay width
for deuterium is taken to be $\Gamma(d)=1-10^3$ eV. 

Figure \ref{fig:timecompare} shows that the lifetimes for stars using
the CNO cycle are relatively insensitive to the metallicity and are
comparable to the lifetimes expected for stars in our universe (for a
given mass).  The lifetimes for stars using the triple nucleon process
are systematically shorter, and vary significantly with the decay
width $\Gamma(d)$. As the decay width increases, the lifetime of
unstable deuterium decreases, and the stars have to contract further
in order to achieve sustained nuclear fusion. This trend results in
hotter stars that are brighter for a given mass and hence
shorter-lived. In addition, stars operating through the triple nucleon
process have central densities and temperatures comparable to those
for helium burning stars in our universe.  As a result, these stars
tend to burn their helium into carbon (through the usual triple alpha
process) at the same time they burn hydrogen into helium (through the
triple nucleon process). These stars thus produce nuclei with $A>5$
(especially carbon) and leave the main-sequence sooner. The smallest
stars (with $\Gamma(d)=1$ eV) live up to $\sim1$ Gyr, perhaps long
enough for accompanying planets to become habitable, whereas stars
with larger widths have shorter lifetimes. Even if the hydrogen
burning timescales are too short for life, however, the first
generation of stars can produce carbon through the triple nucleon and
triple alpha processes, so that later stellar generations will have
enough carbon to operate through the CNO cycle.

\section{Big Bang Nucleosynthesis without Stable Deuterium} 
\label{sec:bbn}  

The discussion thus far has assumed that a universe without stable
deuterium will emerge from its early epochs with essentially no
elements other than hydrogen. This assumption stands in contrast to
the case of our universe, which processes about one fourth of its mass
into helium, along with small (but nonzero) abundances of other light
nuclei. The triple-nucleon reaction considered for stars could in
principle instigate the production of light nuclei during the BBN
epoch. This section considers the early phases of evolution for
universes without stable deuterium.

The triple-nucleon reaction relies on the presence of free neutrons.
In the early universe, the plasma of charged leptons and neutrinos
keeps the neutron-to-proton ratio in weak equilibrium, providing a sea
of free neutrons for big bang nucleosynthesis.  In our universe,
the $n(p,\gamma)d$ and $d(p,\gamma)\heiii$ reaction chain is primarily
responsible for $^3$He production.  We have taken the {\sl\small BURST} 
code \cite{GFKP-5pts:2014mn} and substituted the individual deuterium
reactions with the triple-nucleon reaction.  In addition, we
eliminated the deuterium isotope and all associated nuclear reactions
from the network.  Such a procedure does not preserve unitarity within
the network (see Ref.\ \cite{2014EPJWC..6900003P} for a discussion of
unitarity in BBN).

\begin{figure}
\begin{center}
\includegraphics[scale=0.67]{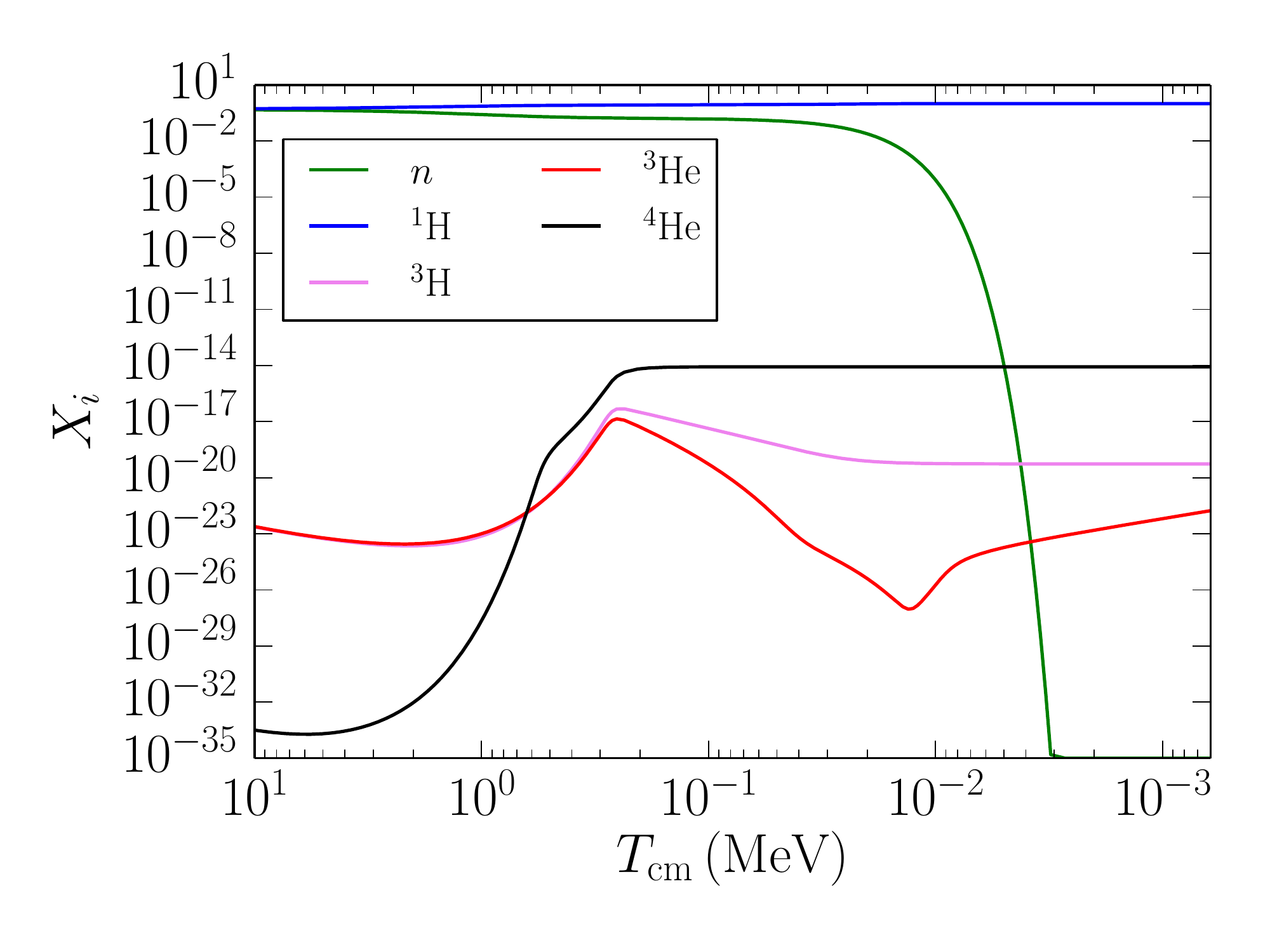}
\end{center}
\caption{\label{fig:BBN} 
Time evolution of light element abundances (mass fractions) during 
Big Bang Nucleosynthesis for a universe without stable deuterium. 
The horizontal axis shows the background temperature $\tcm$ of the
universe as it decreases with increasing time.  The model employed
here uses the triple-nucleon reaction with $\Gamma(d)=1.0\,{\rm eV}$
to synthesize $^3$He from two free protons and one free neutron. The
largest mass fraction with $A>1$ is $^4$He, which reaches a value 
of only $\sim10^{-14}$. Mass fractions for lithium and beryllium 
are below the vertical scale of the plot. }
\end{figure}

Figure \ref{fig:BBN} shows the evolution of the mass fractions of
neutrons, protons (denoted $^1{\rm H}$), ${}^3{\rm H}$, $^3$He, and
$^4$He, with decreasing co-moving temperature parameter $\tcm$
(inversely related to scale factor) during BBN in a universe without
stable deuterium.  The values for the baryon-to-photon ratio and mean
neutron lifetime are identical to those in our universe.  We choose a
value of $\Gamma(d)=1.0\,{\rm eV}$ for the decay width of deuterium.
For high temperatures $\tcm\gtrsim600\,{\rm keV}$, the $\hiii$,
$^3$He, and $^4$He abundances remain in NSE.  At $\tcm\simeq600$ keV,
the $^4$He abundance begins to depart from NSE, as evidenced by the
shoulder in the $^4$He curve in Figure\ \ref{fig:BBN}.  The isotopes
$\hiii$ and $^3$He remain in equilibrium longer, until $\tcm$ reaches
roughly $200\,{\rm keV}$.  Once the abundances depart from NSE, there
is little increase in $^4$He, a strict decrease in $\hiii$, and a
provisional decrease in $^3$He.  These trends result from the
triple-nucleon reaction being much too slow compared to the Hubble
expansion rate to keep producing $^3$He for eventual incorporation
into a $^4$He nucleus.  In our universe, there is no such restriction.
A sea of stable deuterium remains available (due to NSE) until the
temperature falls well below $\tcm=100\,{\rm keV}$, when the deuterium
undergoes out-of-equilibrium synthesis into $^4$He.  

We did not include the reaction $n + n + p \leftrightarrow {}^3{\rm H}
+ \gamma$.  In our universe, $d(p,\gamma)\heiii$ is the dominant
channel for synthesis to nuclei with $A=3$, so including the
$2n(p,\gamma){}^3{\rm H}$ reaction in the BBN network would only
modestly increase the final abundance of $^4$He in Figure\ \ref{fig:BBN}.
The abundance of $\hiii$ is larger than that of $^3$He at the lowest
temperatures in the plot.  This is unusual as $^3$He is the stable
nuclear configuration for $A=3$, and so we would expect a larger
abundance of $^3$He compared to $\hiii$ due to equilibrium arguments.
NSE is no longer obtained for $\tcm\lesssim200\,{\rm keV}$, and so the
abundances of $\hiii$ and $^3$He evolve with the out-of-equilibrium
nuclear rates.  Transmutation of $\hiii$ has three pathways:
incorporation into $^4$He; beta-decay into $^3$He; or photo-dissociation
into two neutrons and a proton.  First, $^4$He synthesis with $\hiii$
occurs through the $\hiii(p,\gamma)^4{\rm He}$ reaction.  This reaction is
slower than the principal pathway for $^4$He synthesis with $^3$He,
namely $\heiii(\heiii,2p)^4{\rm He}$.  Second, $\hiii$ has a lifetime of
$\sim10\,{\rm years}$, and so the associated rate is much smaller than
the Hubble rate until later times.  The rise in $^3$He at
$\tcm\sim10\,{\rm keV}$ is due to the decay of $\hiii$.  Finally, the
fact that deuterium, and its associated nuclear reactions, are absent
in our BBN scenario, eliminates the pathway for $\hiii$ destruction
through the reverse channel of $d(n,\gamma)\hiii$.  $\hiii$ now must
transmute directly into three nucleons, which would require a larger
energy photon.  However, this last point is moot as we did not include
a $\hiii$ version of the triple-nucleon reaction in our network.  All
told, the three pathways cannot transmute $\hiii$ to $^3$He within the
span of time plotted on Figure\ \ref{fig:BBN}.

The end result of these calculations is that BBN is nearly inert to
the triple-nucleon reaction. The results in Figure \ref{fig:BBN} are
presented for the particular decay width $\Gamma(d)=1.0\,{\rm eV}$,
but the value of $\Gamma(d)$ would have to be much smaller, so that
deuterium is far more long-lived, in order for heavier nuclei to
develop substantial abundances during the BBN epoch.  As a result, if
deuterium is unstable, the universe emerges from its early epochs with
an almost pure hydrogen composition. In order for such a universe to
become habitable, most nucleosynthesis must take place later in
stellar cores (or other stellar environments). 

\section{Conclusion} 
\label{sec:conclude} 

This paper considers stellar evolution in universes in which deuterium
has no stable bound state. Previous authors have argued that universes
in this class would not allow for nucleosynthesis to take place due to
the lack of a stable intermediate nucleus between hydrogen and helium. 
In contrast, we show that stars in deuterium-free universes can
provide both energy and nucleosynthesis, and thus allow such universes
to be potentially habitable.

\subsection{Summary of Results} 
\label{sec:results} 

The results of this work, as summarized below, indicate that a number
of different stellar processes can provide both luminosity and
nucleosynthesis in the absence of stable deuterium:

[1] Stars can provide enough energy for habitability, over
sufficiently long time scales, through the action of gravitational
contraction. The time scale over which the luminosity is relatively
large has a maximum value of a few Gyr for a stellar mass
$M_\ast\sim0.5M_\odot$ (see Figure \ref{fig:habtime}). Somewhat larger
stars have shorter lifetimes but their luminosity is more constant in
time (Figure \ref{fig:lumtime}). Stars with initial masses below the
Chandrasekhar limit can be supported by electron degeneracy pressure
at the end of the gravitational contraction phase. These stars end
their lives as white dwarfs. Stars with initial masses above the
Chandrasekhar limit cannot be supported by degeneracy pressure and
collapse at the end of their lives. These stars thus experience an
explosive end state analogous to Type Ia supernovae.  These explosions
can provide heavy elements for subsequent stellar generations (Section
\ref{sec:explode}).

[2] Under sufficiently hot and dense conditions, nucleosynthesis can
take place through a class of triple-nucleon reactions (Section
\ref{sec:ppp}).  Even with no stable state, a small population of
deuterium nuclei will be present in stellar cores, although the
forward reaction rates are not fast enough to reach nuclear
statistical equilibrium (Section \ref{sec:nse}). This population of
deuterium can interact with protons to produce helium, which is
assumed to be stable. The deuterium nuclei also decay into free
neutrons, which have larger reaction rates, and allow for a chain of
nuclear reactions that produce helium (Section \ref{sec:trinuke}).
The resulting triple-nucleon process is roughly analogous to the
triple-alpha reaction through which helium is synthesized into carbon
in intermediate mass stars. The central cores of ordinary stars can
reach the temperatures and densities required for the triple-nucleon
process to take place and thereby generate robust stellar luminosities
(Section \ref{sec:eresults}). Compared to stars burning hydrogen
through conventional reactions, these stars are somewhat brighter and
have higher surface temperatures (see Figure \ref{fig:NSE_vs_3nuc}).
The triple-nucleon process can also occur during the final collapse
phases of stars evolving through gravitational contraction only (see
Figure \ref{fig:temptime} and Section \ref{sec:explode}).

[3] Stars can also burn hydrogen through the CNO cycle. This process,
which dominates energy production in stars more massive than the Sun,
does not require deuterium, but does require a nonzero abundance of
carbon. The CNO cycle can operate with only trace amounts of carbon,
specifically, with metallicities as low as $Z = 10^{-10}$ for solar
type stars (Section \ref{sec:cno}). This process can thus operate as
long as a few stars per galaxy experience explosive nucleosynthesis
(Section \ref{sec:gravity}) in a previous stellar generation and/or if
nucleosynthesis takes places via the triple-nucleon reaction (Section
\ref{sec:ppp}). The resulting stars have properties similar to those
in our universe, with a relatively normal main-sequence (Figure
\ref{fig:cnozams}) and long lifetimes (Figure \ref{fig:timecompare}).

[4] The nuclear processes outlined above are not mutually exclusive,
so that stars can derive energy through all of these channels over the
course of their lifetimes (Section \ref{sec:combine}).  In general,
stars in universes without stable deuterium will begin their evolution
with gravitational contraction. The central temperature increases with
time, and eventually the stellar core becomes hot and dense enough for
the triple-nucleon reactions to ignite and arrest further contraction.
The conditions required for the triple-nucleon reaction to operate are
similar to those required for helium burning, and hence carbon
production, in ordinary stars. As a result, some of the helium
produced via the triple-nucleon reaction will be burned into carbon
through the triple-alpha process. When enough carbon builds up in the
stellar core, energy generation through the CNO cycle can dominate
(Figure \ref{fig:3nuc_set}). Note that the CNO cycle operates at lower
central temperatures than the triple nucleon process. For subsequent
stellar generations that are formed with nonzero carbon abundance, 
stars will burn their hydrogen through the CNO cycle before reaching
the conditions where the triple nucleon process can take place. 

[5] Although stellar nucleosynthesis can take place through alternate
channels, Big Bang Nucleosynthesis is not effective in universes with
no stable deuterium (Section \ref{sec:bbn}). Compared to stellar
conditions, the density of the universe is much lower during BBN, so
that triple-nucleon reactions are highly suppressed. As a result, the
universe ends up with only trace amounts of $^4$He ($X_4\sim10^{-14}$)
and even lower abundances of other nuclei (Figure \ref{fig:BBN}). 

\subsection{Discussion} 
\label{sec:discuss} 

In our universe, the role played by stars varies with stellar mass.
Low mass stars dominate the mass budget and live long enough to serve
as hosts for habitable planets.  High mass stars are rarer, but they
dominate the generation of energy and the production of heavy elements
necessary for life.  In the alternate scenarios considered here, where
deuterium has no stable state, high mass and low mass stars play
analogous roles.  

In the extreme limit where no (steady) nuclear processing takes place,
a sharp division arises at the Chandrasekhar mass (where
$\mchan\approx5.6M_\odot$ for stars composed of hydrogen).  Stars of
all masses contract as they evolve and produce energy through the loss
of gravitational potential energy.  Stars with $M_\ast>\mchan$ evolve
too rapidly to host habitable planets. They are also too heavy to be
supported by degeneracy pressure and eventually experience
catastrophic collapse, so that their central temperatures and
densities increase to enormous values. Under these extreme conditions,
the absence of stable deuterium no longer enforces a bottleneck on
nuclear processing and heavier elements can be produced.  In contrast,
stars with $M_\ast<\mchan$ contract until they reach maximum
temperatures and densities, where the values depend on stellar mass.
These stars have highly diminished capacity for producing heavy
elements, but they retain significant luminosities (comparable to
stars in our universe) over long spans of time (up to billions of
years).

In the case where the triple-nucleon process can operate, stars evolve
much like those in our universe. The high mass stars are brighter, and
hotter, and live for shorter spans of time. Stars of lower mass can
live long enough to support habitable planets. Compared to stars in
our universe, stars without stable deuterium must have a higher mass
in order to achieve the same level of nuclear burning, and they have a
shorter lifetime for a given mass.  More specifically, the conditions
required for hydrogen burning via the triple-nucleon process are
comparable to those required for helium burning (carbon production)
via the triple-alpha process. As a result, the minimum mass for
sustained hydrogen burning, the brown dwarf limit, will occur at a
higher mass than in our universe.\footnote{Note that the mass scale  
for the brown dwarf limit will depend on the parameters that 
characterize the instability of deuterium, e.g., the decay width
$\Gamma(d)$ from Section \ref{sec:trinuke}.} At the other end of the
mass spectrum, however, the maximum stellar mass is set by the onset
of radiation pressure domination \cite{phil} and is not expected to
change. The allowed range of stellar masses thus decreases.

This paper has explored a number of nuclear processes that allow stars
to generate energy and produce heavy elements in the absence of stable
deuterium, including explosive nucleosynthesis, triple-nucleon
reactions, and the CNO cycle.  The relative importance of these
processes determines how stellar evolution ultimately occurs and
depends on the mass difference between deuterium and its constituent
particles. As deuterium nuclei become less stable, the nuclear
processes considered in this paper become less efficient. For example,
if deuterium nuclei are farther from stability, the radiative decay
width $\Gamma(d)$ is larger, and stars require higher central
temperatures for the triple-nucleon process to operate. In addition to
making it more difficult for stars to burn hydrogen on the
main-sequence, greater instability of deuterium reduces the efficacy
of explosive nucleosynthesis. Although the CNO cycle operates without
deuterium, it relies on carbon having a nonzero abundance, which in
turn depends on the aforementioned nuclear processes involving
deuterium. This paper argues that universes can remain habitable
without stable deuterium.  However, if deuterium nuclei were to become
sufficiently unstable, the nuclear processes considered herein would
become so inefficient that the universe would still end up lifeless. 
Such limits should be explored further in future work.

This paper assumes that deuterium is unstable but helium isotopes are
bound. As a result, the universes under consideration here are roughly
similar to our own. In order to quantify this issue, one can consider
the possible parameter space for the Standard Model of particle
physics where the quark masses are allowed to vary
\cite{hogan,damour,jaffe}.  For particle physics models with two light
quark species with different charges, like our own universe, such
worlds will support analogs of protons, neutrons, and the
corresponding composite nuclei. In order for hydrogen to have a stable
isotope, the mass difference between the nucleons must satisfy the
constraint $\mpro-\mneut<7.97$ MeV, whereas the requirement for a
stable carbon nucleus implies $\mneut-\mpro<14.77$ MeV
\cite{jaffe}. For comparison, the electron mass $m_e=0.511$ MeV and
the mass difference in our universe $\mneut-\mpro\approx1.29$
MeV. According to this analysis, the quark masses must conspire to
produce differences in nucleon masses that fall within a range of
about 23 MeV. It is significant that all of these energies are small
compared to the benchmark scale provided by the Higgs vacuum
expectation value, which falls at $\sim246$ GeV. The allowed window
for nucleon mass differences to support stable nuclei (23 MeV) is thus
relatively narrow compared to the Higgs scale (246 GeV), but still
wider than the mass difference in our universe (1.3 MeV). In order to
assess whether this allowed range for habitable universes is large or
small, one needs the underlying probability distribution for the
possible quark masses and other parameters of the Standard Model.
These distributions are currently unavailable, but will hopefully
become better understood in the future.

The results of this paper expand the range of potentially habitable
universes to include some fraction of those without stable deuterium.
Previous work \cite{barnes2015} has already shown that stars can
operate in universes where the strong force is more effective so that
diprotons are bound.  Variations in the strong force also influence
the location of the carbon resonance that allows the triple alpha
reaction to efficiently produce carbon in our universe. In such
universes, however, $^8$Be can (sometimes) be stable, so that carbon
can be produced via non-resonant reactions \cite{adgroh}. Still other
work shows that stars can operate --- with sufficiently long lifetimes
and hot surface temperatures --- over a wide range of values for the
fine structure constant, the gravitational constant, and nuclear
reaction rates \cite{adams2008,adamsnew}. Taken together, these
results indicate that stellar evolution has many different pathways,
and that stars are robust enough to provide both energy and
nucleosynthesis over a wide range of possible universes.

\bigskip 
\noindent
{\bf Acknowledgments:} We would like to thank George Fuller, David
Garfinkle, and Mark Paris for useful discussions. We also thank an
anonymous referee for many useful comments that improved the paper.
This work was supported by the JT Foundation through Grant ID55112:
{\sl Astrophysical Structures in other Universes}, and by the
University of Michigan. Computational resources and services were
provided by Advanced Research Computing at the University of Michigan.

\end{document}